\documentclass[twocolumn]{aastex631}
\usepackage{enumitem}

\newcommand{\one}{$\rm (1-0)$}
\newcommand{\two}{$\rm (2-1)$}

\newcommand{\s}{\,{\rm s}}

\newcommand{\m}{\,{\rm m}}

\newcommand{\mm}{\,{\rm mm}}
\newcommand{\mum}{\,{\mu {\rm m}}}

\newcommand{\as}{^{\prime\prime}}

\newcommand{\bdm}{\begin{displaymath}}
\newcommand{\edm}{\end{displaymath}}
\newcommand{\beq}{\begin{equation}}
\newcommand{\eeq}{\end{equation}}
\newcommand{\bit}{\begin{itemize}}
\newcommand{\eit}{\end{itemize}}
\newcommand{\ben}{\begin{enumerate}}
\newcommand{\een}{\end{enumerate}}
\newcommand{\bfi}{\begin{figure}[htb]}
\newcommand{\bpfi}{\begin{figure}[p]}

\begin{document}

\title{In-orbit Performance of the Near-Infrared Spectrograph NIRSpec on the James Webb Space Telescope}

\author[0000-0002-5666-7782]{T. B\"oker}
\affiliation{European Space Agency, c/o STScI, 3700 San Martin Drive, Baltimore, MD 21218, USA}

\author[0000-0002-6881-0574]{T. L. Beck}
\affiliation{Space Telescope Science Institute, 3700 San Martin Drive, Baltimore, MD 21218, USA}

\author[0000-0001-7058-1726]{S. M. Birkmann}
\affiliation{European Space Agency, c/o STScI, 3700 San Martin Drive, Baltimore, MD 21218, USA}

\author[0000-0002-9262-7155]{G. Giardino}
\affiliation{ATG Europe for the European Space Agency, ESTEC, Noordwijk, The Netherlands}

\author[0000-0002-4834-369X]{C. Keyes}
\affiliation{Space Telescope Science Institute, 3700 San Martin Drive, Baltimore, MD 21218, USA}

\author[0000-0002-5320-2568]{N. Kumari}
\affiliation{AURA for the European Space Agency,
Space Telescope Science Institute, 3700 San Martin Drive, Baltimore, MD 21218, USA}

\author{J. Muzerolle}
\affiliation{Space Telescope Science Institute, 3700 San Martin Drive, Baltimore, MD 21218, USA}

\author[0000-0002-7028-5588]{T. Rawle}
\affiliation{European Space Agency, c/o STScI, 3700 San Martin Drive, Baltimore, MD 21218, USA}

\author[0000-0002-6091-7924]{P. Zeidler} 
\affiliation{AURA for the European Space Agency,
Space Telescope Science Institute, 3700 San Martin Drive, Baltimore, MD 21218, USA}

\author{Y. Abul-Huda}
\affiliation{Space Telescope Science Institute, 3700 San Martin Drive, Baltimore, MD 21218, USA}

\author[0000-0003-2896-4138]{C. Alves de Oliveira}
\affiliation{European Space Agency, ESAC, Villanueva de la Cañada, E-28692 Madrid, Spain}

\author[0000-0001-7997-1640]{S. Arribas}
\affiliation{Centro de Astrobiolog\'ia (CAB), CSIC–INTA, Cra. de Ajalvir Km.~4, 28850- Torrej\'on de Ardoz, Madrid, Spain}

\author[0000-0002-7722-6900]{K. Bechtold}
\affiliation{Space Telescope Science Institute, 3700 San Martin Drive, Baltimore, MD 21218, USA}

\author[0000-0003-0883-2226]{R. Bhatawdekar}
\affiliation{European Space Agency, ESTEC, Keplerlaan 1, 2201 AZ Noordwijk, The Netherlands}

\author[0000-0001-8470-7094]{N. Bonaventura}
\affiliation{Cosmic Dawn Center (DAWN), Niels Bohr Institute, University of Copenhagen, Jagtvej 128, DK-2200, Denmark}

\author{A. J. Bunker}
\affiliation{Department of Physics, University of Oxford, Denys Wilkinson Building, Keble Road, Oxford, OX1 3RH, UK}

\author[0000-0002-0450-7306]{A. J. Cameron}
\affiliation{Department of Physics, University of Oxford, Denys Wilkinson Building, Keble Road, Oxford, OX1 3RH, UK}

\author[0000-0002-6719-380X]{S. Carniani}
\affiliation{Scuola Normale Superiore, Piazza dei Cavalieri 7, I-56126 Pisa, Italy}

\author[0000-0003-3458-2275]{S. Charlot}
\affiliation{Sorbonne Universit\'e, UPMC-CNRS, UMR7095, Institut d'Astrophysique de Paris, F-75014 Paris, France}

\author[0000-0002-2678-2560]{M. Curti}
\affiliation{Cavendish Laboratory, University of Cambridge, 19 J. J. Thomson Ave., Cambridge CB3 0HE, UK}
\affiliation{Kavli Institute for Cosmology, University of Cambridge, Madingley Road, Cambridge CB3 0HA, UK}

\author[0000-0001-9513-1449]{N. Espinoza}
\affiliation{Space Telescope Science Institute, 3700 San Martin Drive, Baltimore, MD 21218, USA}
\affiliation{Department of Physics \& Astronomy, Johns Hopkins University, Baltimore, MD 21218, USA}

\author[0000-0001-8895-0606]{P. Ferruit}
\affiliation{European Space Agency, ESAC, Villanueva de la Cañada, E-28692 Madrid, Spain}

\author[0000-0002-8871-3027]{M. Franx}
\affiliation{Leiden Observatory, Leiden University, PO Box 9513, 2300RA Leiden, The Netherlands}

\author[0000-0002-6780-2441]{P. Jakobsen}
\affiliation{Cosmic Dawn Center (DAWN), Niels Bohr Institute, University of Copenhagen, Jagtvej 128, DK-2200, Denmark}

\author{D. Karakla}
\affiliation{Space Telescope Science Institute, 3700 San Martin Drive, Baltimore, MD 21218, USA}

\author[0000-0003-1016-9283]{ M.~L\'{o}pez-Caniego}
\affiliation{Aurora Technology for the European Space Agency, Villanueva de la Ca\~nada, E-28692 Madrid, Spain.}
\affiliation{Universidad Europea de Madrid, 28670, Madrid, Spain.}

\author[0000-0002-4034-0080]{N. L\"utzgendorf}
\affiliation{European Space Agency, c/o STScI, 3700 San Martin Drive, Baltimore, MD 21218, USA}

\author[0000-0002-4985-3819]{R. Maiolino}
\affiliation{Cavendish Laboratory, University of Cambridge, 19 J. J. Thomson Ave., Cambridge CB3 0HE, UK}
\affiliation{Kavli Institute for Cosmology, University of Cambridge, Madingley Road, Cambridge CB3 0HA, UK}

\author[0000-0003-0192-6887]{E. Manjavacas}
\affiliation{AURA for the European Space Agency,
Space Telescope Science Institute, 3700 San Martin Drive, Baltimore, MD 21218, USA}

\author[0000-0001-5788-5258]{A. P. Marston}
\affiliation{European Space Agency, ESAC, Villanueva de la Cañada, E-28692 Madrid, Spain}

\author{S. H. Moseley}
\affiliation{Quantum Circuits, Inc., New Haven, Connecticut, USA}

\author[0000-0002-3471-981X]{P. Ogle}
\affiliation{Space Telescope Science Institute, 3700 San Martin Drive, Baltimore, MD 21218, USA}

\author[0000-0002-0362-5941]{M. Perna}
\affiliation{Centro de Astrobiolog\'ia (CAB), CSIC–INTA, Cra. de Ajalvir Km.~4, 28850- Torrej\'on de Ardoz, Madrid, Spain}

\author[0000-0003-2314-3453]{M. Pe\~na-Guerrero}
\affiliation{Space Telescope Science Institute, 3700 San Martin Drive, Baltimore, MD 21218, USA}

\author{N. Pirzkal}
\affiliation{AURA for the European Space Agency,
Space Telescope Science Institute, 3700 San Martin Drive, Baltimore, MD 21218, USA}

\author[0000-0002-2509-3878]{R. Plesha}
\affiliation{Space Telescope Science Institute, 3700 San Martin Drive, Baltimore, MD 21218, USA}

\author[0000-0001-7617-5665]{C. R. Proffitt}
\affiliation{Space Telescope Science Institute, 3700 San Martin Drive, Baltimore, MD 21218, USA}

\author[0000-0003-2662-6821]{B. J. Rauscher} 
\affiliation{NASA Goddard Space Flight Center, Observational Cosmology Laboratory, Greenbelt, USA}

\author{H.-W. Rix }
\affiliation{Max-Planck Institute for Astronomy, Königstuhl 17, 69117 Heidelberg, Germany}

\author[0000-0001-5171-3930]{B. Rodr\'iguez del Pino}
\affiliation{Centro de Astrobiolog\'ia (CAB), CSIC–INTA, Cra. de Ajalvir Km.~4, 28850- Torrej\'on de Ardoz, Madrid, Spain}

\author{Z. Rustamkulov}
\affil{Department of Earth \& Planetary Sciences, Johns Hopkins University, Baltimore, MD 21218, USA}

\author[0000-0003-2954-7643]{E. Sabbi}
\affiliation{Space Telescope Science Institute, 3700 San Martin Drive, Baltimore, MD 21218, USA}

\author[0000-0001-6050-7645]{D. K. Sing}
\affil{Department of Physics \& Astronomy, Johns Hopkins University, Baltimore, MD 21218, USA}
\affil{Department of Earth \& Planetary Sciences, Johns Hopkins University, Baltimore, MD 21218, USA}

\author{M. Sirianni}
\affiliation{European Space Agency, c/o STScI, 3700 San Martin Drive, Baltimore, MD 21218, USA}

\author{M. te Plate}
\affiliation{European Space Agency, c/o STScI, 3700 San Martin Drive, Baltimore, MD 21218, USA}

\author[0000-0001-7130-2880]{L. \'Ubeda}
\affiliation{Space Telescope Science Institute, 3700 San Martin Drive, Baltimore, MD 21218, USA}

\author[0000-0002-6570-4776]{G. M. Wahlgren}
\affiliation{Space Telescope Science Institute, 3700 San Martin Drive, Baltimore, MD 21218, USA}

\author{E. Wislowski}
\affiliation{Space Telescope Science Institute, 3700 San Martin Drive, Baltimore, MD 21218, USA}

\author{R. Wu}
\affiliation{Space Telescope Science Institute, 3700 San Martin Drive, Baltimore, MD 21218, USA}

\author[0000-0002-4201-7367]{Chris J. Willott}
\affil{NRC Herzberg, 5071 West Saanich Rd, Victoria, BC V9E 2E7, Canada}

\begin{abstract}
The Near-Infrared Spectrograph (NIRSpec) is one of the four focal plane instruments on the James Webb Space Telescope. In this paper, we summarize the in-orbit performance of NIRSpec, as derived from data collected during its commissioning campaign and the first few months of nominal science operations. More specifically, we discuss the performance of some critical hardware components such as the two NIRSpec Hawaii-2RG (H2RG) detectors, wheel mechanisms, and the micro-shutter array. We also summarize the accuracy of the two target acquisition procedures used to accurately place science targets into the slit apertures, discuss the current status of the spectrophotometric and wavelength calibration of NIRSpec spectra, and provide the 'as measured' sensitivity in all NIRSpec science modes. Finally, we point out a few important considerations for the preparation of NIRSpec science programs.
\end{abstract}


\keywords{Instrumentation: spectrographs - Space vehicles: instruments}

\section{Introduction and Background}\label{sec:intro}
Many, if not most, of the science goals for the James Webb Space Telescope (JWST) rely on the ability to acquire high-quality near-infrared (NIR) spectra of astronomical targets with a wide range of luminosities, from the faintest galaxies at high redshift to the bright host stars of nearby exoplanets. The NearInfraRed Spectrograph (NIRSpec) onboard JWST offers  the necessary versatility to enable observations of such a wide range of targets, thanks to a variety of sophisticated observing modes, some of which are available for the first time in space and at near-infrared wavelengths (e.g. multi-object and integral field spectroscopy). 

In this paper, we summarize the status of the NIRSpec performance characterization, using mostly data acquired during the JWST commissioning campaign. The analysis of these data sets (which in many cases have since been complemented by additional data acquired during Cycle 1) is still ongoing, and many performance-related measurements and products are preliminary. Therefore, this paper can only provide a snapshot of the performance assessment as of mid-Oct. 2022.

\section{Overview of the NIRSpec Instrument}
\begin{figure*}[htb]
    \centering
    \includegraphics[width=0.96\columnwidth]{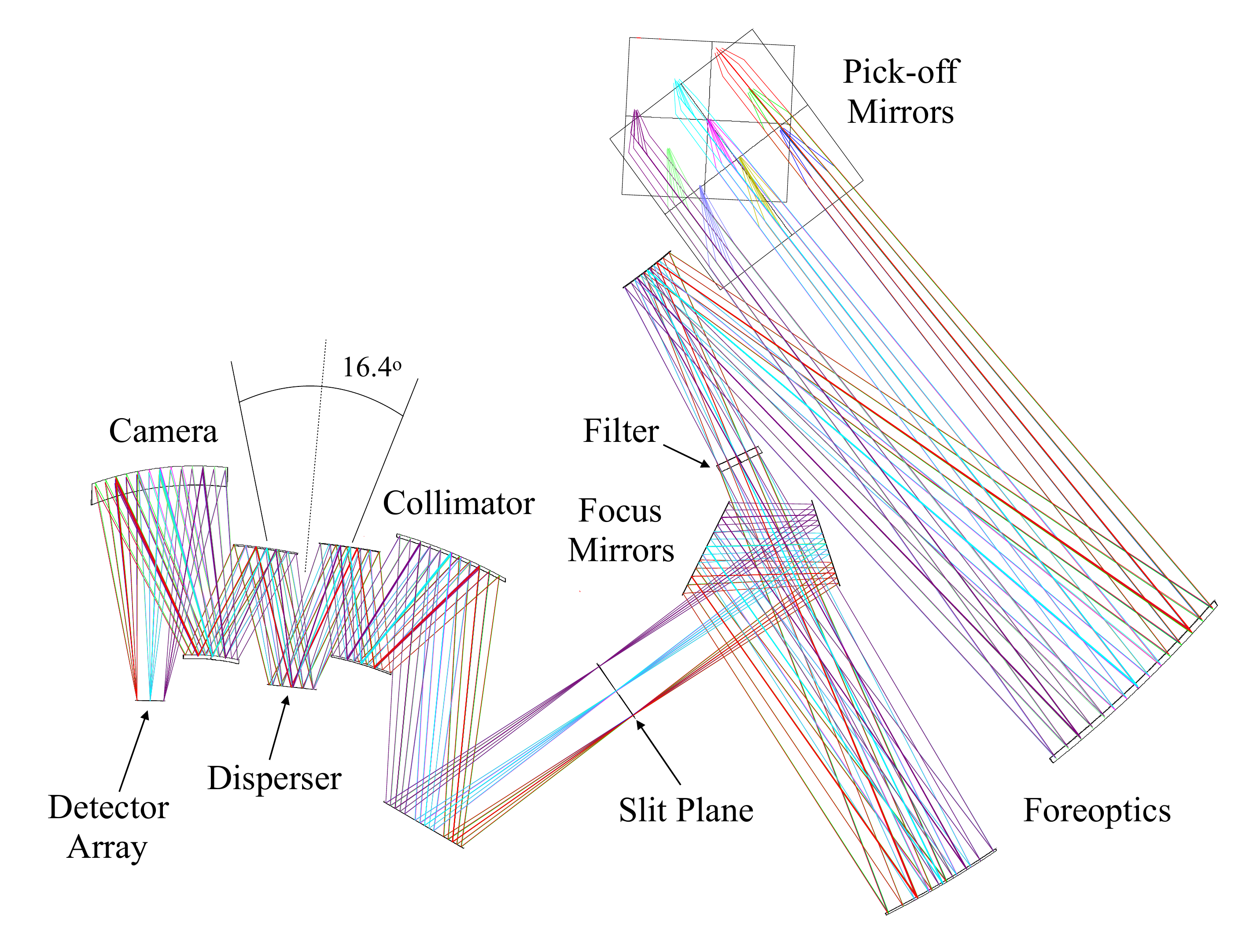}
    \includegraphics[width=0.96\columnwidth]{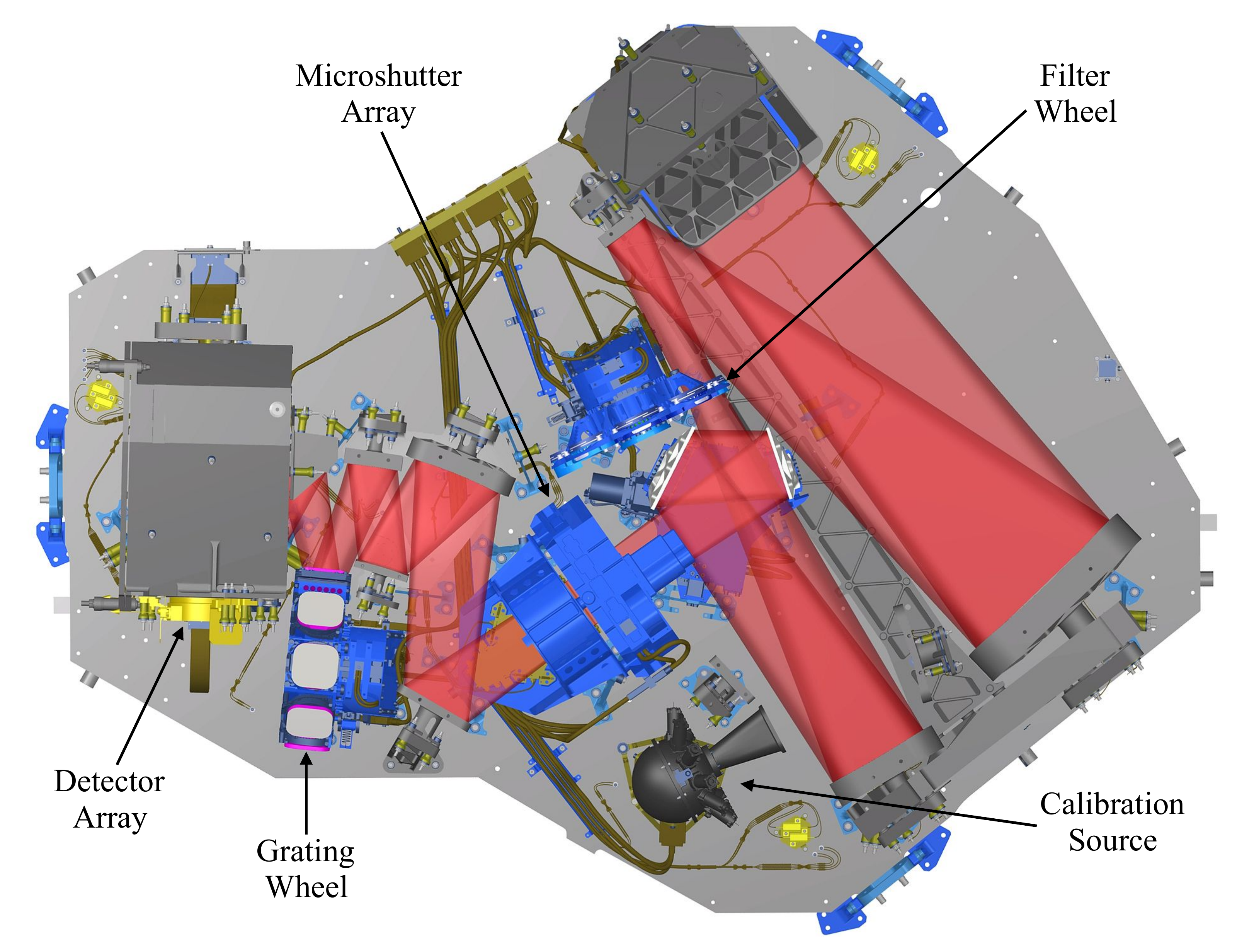}
\caption{Left: The optical path through the NIRSpec instrument. The dispersion direction is out of the page. Right: CAD rendering of NIRSpec with its major mechanisms identified. The dimensions of the NIRSpec optical assembly are 1.9\,m $\times$ 1.4\,m $\times$ 0.7\,m, with a total mass of 196\,kg.}
    \label{fig:layout}
\end{figure*}

The design of and scientific motivation for the NIRSpec instrument have been extensively discussed in \cite{jakobsen22}. For convenience, we show here again the optical design of the NIRSpec instrument, together with the layout of the mechanical assembly in Figure\,\ref{fig:layout}.

In brief, NIRSpec is designed to be capable of performing both single- and multi-object spectroscopic observations over the $0.6-5.3\mum$ NIR wavelength range at three spectral resolutions that cover a range of science applications: a low-resolution ($R\simeq$ 30--330) mode intended for obtaining exploratory continuum spectra and redshifts of remote galaxies; an intermediate resolution ($R\simeq1000$) mode for accurately measuring atomic and molecular emission lines, and a higher resolution ($R\simeq2700$, i.e. $\simeq 110$\,km/s) mode primarily for kinematic studies using these emission lines. Table\,\ref{tab:modes} summarizes the various NIRSpec modes and their typical science applications, together with their slit apertures and wavelength coverage, and available spectral resolution.
More detailed descriptions of the NIRSpec observing modes can be found in \cite{ferruit22} for the Multi-Object Spectroscopy (MOS) mode, \cite{boeker22} for the Integral-Field Spectroscopy (IFS) mode, and \cite{birkmann22a} for the Bright Object Transit Spectroscopy (BOTS) mode.

\begin{table}[htb]
\caption{NIRSpec instrument modes. }\label{tab:modes}
\centering
    \includegraphics[width=0.96\columnwidth]{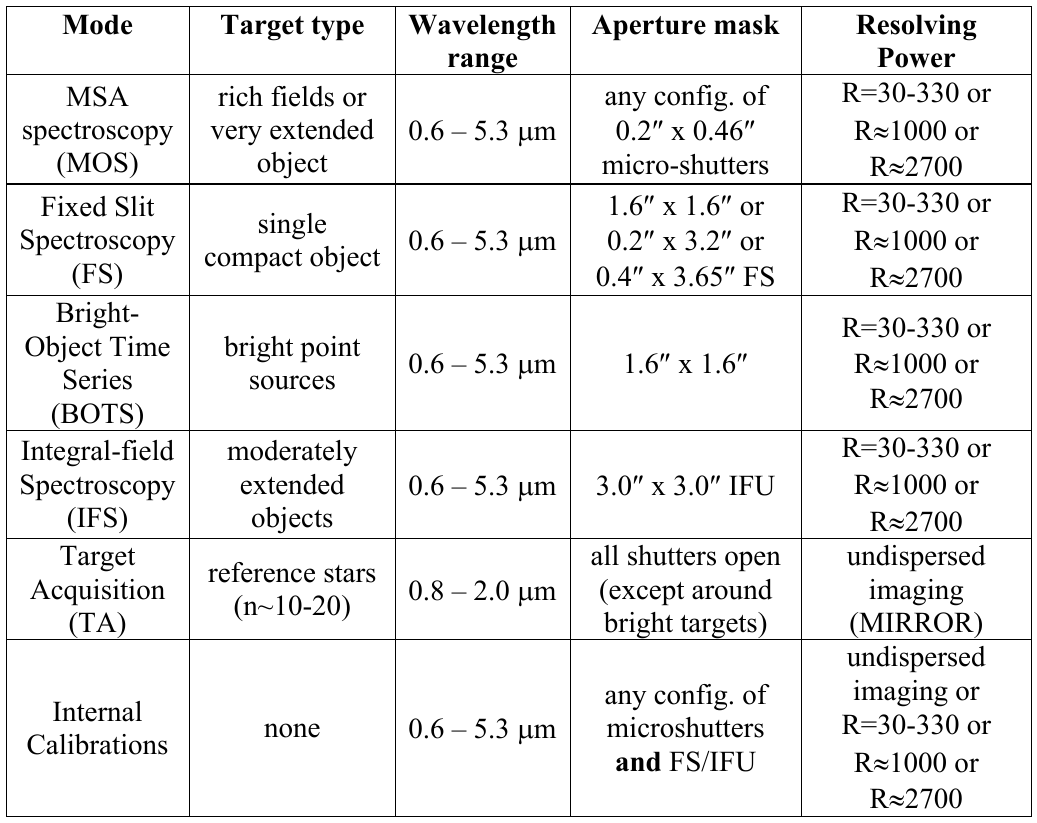}
\end{table}

\section{The NIRSpec Commissioning Campaign}
As discussed in more detail in \cite{boeker22spie}, the NIRSpec commissioning campaign successfully concluded 185 days after launch, when the last of its four observing modes was formally declared `ready for science'. A total of 33 NIRSpec commissioning activity requests (CARs) were required to activate the instrument electronics, to commission the various observing modes, and to obtain a first level of calibration and performance evaluation. While the activities were scattered throughout the six months of JWST commissioning, they had to be executed in a carefully planned sequence in order to derive the various intermediate products required for subsequent, and increasingly more complex, activities. The major milestones were the following:

\begin{enumerate}
  \setlength{\itemsep}{3pt}

\item The successful execution of an onboard script that controls and maintains the temperature of the microshutter quadrants 10-15K above the environment in JWST's Integrated Science Instrument Module (ISIM), until the latter dropped below 140K, in order to avoid condensation of water ice or other volatiles onto the micro-shutters.
    
    \item The NIRSpec Optical Bench Assembly (OBA) reaching sufficiently cold temperatures for the safe operation of the various mechanisms (the magnet arm of the microshutter array (MSA), the wheel mechanisms, and the re-focus mechanism) and the active control of the Focal Plane Assembly (FPA) temperature.
    
    \item The verification of the mechanical and operational performance of these mechanisms and the various lamps in the Calibration Assembly (CAA).
    
    \item The detailed performance characterization of the NIRSpec detectors and the $\approx 250000$ individual shutters of the MSA. The results, which are detailed in \cite{birkmann22spie} and \cite{rawle22}, confirmed the expected in-flight performance, and marked the start of Phase~2 of NIRSpec commissioning, i.e. the ability to obtain science exposures via the onboard scripts.
    
    \item The completion of the optical telescope element (OTE) alignment described in \cite{mcelwain22} enabled the first on-sky NIRSpec observations, in particular the `focus sweep' which was used to verify the optimal positioning of the internal focus adjustment mechanism, and allowed a first assessment of the image quality and total wavefront error across the NIRSpec field of view (FOV). For this, the NIRSpec focus mechanism was 'swept' over a range of $\pm 2000$ motor steps, with an exposure in the F110W filter taken every 400 steps. The sweep range corresponds to a physical mechanism movement of $\pm 1\,\mm$ and a defocus of $\pm 3.6\,\mm$ in the OTE focal plane. 
    
    \item Once the optimal NIRSpec focus was established, undispersed images of a dense star field with known and accurate stellar astrometry were obtained. These observations of the `astrometric reference field' \citep{anderson21} enabled precise measurements of the magnification and distortions of the NIRSpec optical train, which is a crucial ingredient for the parametric instrument model \citep{dorner16,luetzgendorf22}, which underlies the wavelength calibration of all NIRSpec spectra (see Sec.~\ref{subsec:wavecal}), as well as the coordinate transformations between detector, MSA, and sky that are required to accurately place science targets in the various NIRSpec apertures.
    
    \item Once the accuracy of the instrument model was confirmed via a series of successful target acquisitions (TAs), the last few activities of the NIRSpec commissioning campaign obtained on-sky observations of stars and planetary nebulae for flux and wavelength calibration, which enabled the creation of various `reference files' and other products needed for the data reduction pipeline.
    
\end{enumerate}

\section{Observatory Performance}
Because of the narrow slit apertures involved in most of its observing modes, the NIRSpec science performance is critically dependent on a number of telescope and observatory characteristics. Here, we briefly list the most important ones, without discussing them in detail because they are the subject of accompanying papers in this issue.

\subsection{Optical Quality of the OTE}\label{subsec:OTE}
The quality of the image delivered by the OTE is of fundamental importance for all JWST instruments, as it limits the spatial resolution and sensitivity of images. As discussed in other papers in this edition \citep{menzel22,mcelwain22}, the in-orbit optical performance of the OTE is significantly better than expected, with lower wavefront errors and therefore a sharper point spread function (PSF) across the entire FOV \citep{feinberg22}. More specifically, the JWST PSF is diffraction limited already at $1.1\mum$, and reaches a Strehl ratio of 0.9 at $4\mum$ across the FOV.

While this is clearly good news for all JWST instruments, the NIRSpec sensitivity stands to gain the most. This is because a narrower PSF minimizes the `path losses' caused by the physical truncation and subsequent diffraction of the PSF by the narrow apertures of the NIRSpec fixed slits and microshutters. As discussed in \cite{giardino22}, this is the main reason for the fact that the NIRSpec slitlosses are significantly smaller than expected, with peak gains of more than 10\% at the shorter wavelengths.
The resulting benefit to the NIRSpec throughput is discussed in Section\,\ref{subsec:pce}.

\subsection{Cleanliness and thermal backgrounds}
In addition to the low wavefront error, the OTE optics are also free of water ice or other molecular contamination, resulting in a very high telescope throughput across the entire NIRSpec wavelength range. Moreover, the good thermal performance of the sunshield, and the resulting nominal temperatures of the OTE result in thermal backgrounds that are at or below the pre-launch expectations \citep{rigby22b}, also enhancing the sensitivity of NIRSpec observations. Taken together, the outstanding optical and thermal performance of the JWST OTE provides a much better wavefront to the NIRSpec optics than expected, which is reflected in the sensitivity numbers discussed in Section\,\ref{subsec:pce}.

\subsection{Attitude Control System}\label{ACS}
Another important factor for the quality of NIRSpec observations is the pointing accuracy and stability of the JWST Attitude Control System (ACS), which is discussed in detail in \cite{menzel22}. Accurate `blind pointing' is required to ensure that the science target is indeed imaged onto the rather small detector area used to identify the target by the onboard TA algorithms, and that the corrective small-angle maneuvers (SAMs) are not too large. The ability to accurately execute those SAMs, and the absence of any drifts or low-frequency jitter in the telescope pointing is crucial for stable and precise target placement throughout the exposure, which is particularly important for time series observations of transiting exoplanets, as discussed further in Section\,\ref{subsec:TSO}. On all these accounts, the JWST observatory meets or exceeds the pre-launch requirements, which is the foundation for the problem-free execution of the various NIRSpec observing templates, and the outstanding NIRSpec science performance presented in this paper.

\section{NIRSpec Hardware Performance}\label{sec:hardware} 

\subsection{Wheel Mechanisms}\label{subsec:wheels}
NIRSpec is equipped with two wheel mechanisms: i) the Filter Wheel Assembly (FWA) with five long-pass transmission filters (F070LP, F100LP, F170LP, F290LP, and CLEAR) that define the wavelength ranges for the dispersers, two filters used for the TA exposures (F110W and F140X), and one position (OPAQUE) that serves both as an instrument shutter towards the telescope side and a coupling mirror for illumination from the CAA on the instrument side, and ii) the Grating Wheel Assembly (GWA) with the seven NIRSpec dispersers, and a mirror for undispersed imaging (see also Table~\ref{tab:configs}). The functionality of both wheels was verified during commissioning, and their torque profiles were characterized and found to be nominal, i.e. similar to those measured during ground testing, and well within their expected tolerances. 

As discussed in \cite{jakobsen22}, the limited mechanical angular reproducibility of the GWA ball bearings causes small, but significant, variations in the position of NIRSpec spectra on the detector, especially in dispersion direction. Since this has an obvious effect on the wavelength calibration as well as the precision of the TA algorithm, it is important that the actual GWA position of any given NIRSpec exposure can be accurately inferred from telemetry. To this end, the GWA design includes two magneto-resistive position sensors (a.k.a.`tilt sensors') which were extensively tested before launch \citep{demarchi12} to ensure that NIRSpec can indeed achieve the required accuracy of its wavelength calibration.

The tilt sensors have to be calibrated after each cooldown, and their post-launch calibration was one of the important goals of the NIRSpec commissioning campaign. The accuracy of the in-orbit sensor calibration has been discussed in detail by \cite{alves22} : the remaining uncertainties are consistently below 1/10th of a pixel, which is fully in line with expectations and sufficient to ensure a reliable wavelength calibration for all NIRSpec spectra, as demonstrated further in Sec.\,\ref{subsec:wavecal}. The tilt sensor performance and the stability of the wavelength calibration will be monitored with a roughly semi-annual cadence as part of the overall JWST calibration plan, using a combination of stellar spectra and the internal LINE and REF calibration lamps \citep[see][]{jakobsen22}.

\subsection{Microshutter Array}\label{subsec:msa}

The NIRSpec microshutter array (MSA) has performed excellently throughout flight thus far, with no unexpected hardware issues, overall multiplexing remaining very high, and an average success rate of $\sim$96\% for commanding shutters open in science-like patterns \citep{rawle22}. Operability at the level of individual shutters is of course crucial for the MOS mode itself, but less obviously also for the IFS mode, where problems with the shutter array can become a source of parasitic light, either contamination through stuck-open shutters or thermal glow\footnote{while a full in-orbit characterization of the spectral characteristics of MSA shorts is still pending, ground test data have indicated that it is clearly thermal in nature. However, it is not a pure blackbody spectrum, likely because it contains signatures from the MSA coating materials.} emanating from electrical shorts in the circuitry. Our knowledge of this performance is critical to allow mitigation, such as short-masking, and enable users to generate the optimal target set around the inoperable shutter population.

As described in detail by \cite{rawle22}, the marginal increase in inoperable shutters seen during commissioning was in-line with pre-launch estimates. The primary consequence of these trends is that the masking required to mitigate electrical shorts is now the dominant factor affecting the usability of unvignetted shutters (Table\,\ref{tab:msop}). While comparatively fewer shorts have emerged after launch, with approximately one row/column needing to be masked for every 50 shutter array re-configurations compared to one masked per 20--25 during ground testing, shorts have continued to appear throughout the latter stages of commissioning and another two required masking in the first three months of science observations. An added complication apparent for two of the shorts seen since launch (and never during ground testing) was that they produced glow when applying the all-closed shutter configuration for IFS observations. Although successfully mitigated in the usual manner, the contamination before masking affected both MOS and IFS exposures, increasing the overall impact of the shorts.

In the regime that shorts may still occur every few months, there are two operational issues to tackle. First, to ensure that data still meet  scientific objectives, users need to be informed in a timely manner when executed exposures are directly contaminated by a short and also when upcoming observations are affected by newly masked shutters. The former may now apply to both MOS and IFS visits. Second, as the population of shorts continues to grow, each removing several hundred shutters from operation, this will eventually have a noticeable impact on multiplexing. Shorts are believed to be caused by particulate contamination of the complex MSA control electronics, and those particles may shift during reconfiguration, which is the origin of new shorts but also potentially clears the cause of previously detected shorts. Therefore, as discussed by \cite{rawle22}, re-checking whether older shorts still remain may be an avenue to recover previously unusable shutters and maintain the multiplexing of NIRSpec MOS at the current high level.

\begin{table}
\caption{MSA operability report from the end of commissioning, demonstrating that 82.5\% of un-vignetted shutters are available for use as science apertures.}\label{tab:msop}
\centering
\renewcommand{\arraystretch}{1.2} 
\begin{tabular}{|l|rrrr|r|}
\hline
 & Q1 & Q2 & Q3 & Q4 & Total \\
\hline
Total & 62415  &62415&  62415 & 62415  & 249660 \\
Vignetted   &    6119  & 5929 &  6102 &  5874 &   24024    \\
Failed open      &              6  &    3   &  12    &  1     &  22    \\
Short-masked       &        7835  & 5150  & 3466 &  7177 &   23628   \\
Failed closed    &          1569  & 3328 &  5932 &  5064  &  15893 \\
\hline
Total Usable &   46886 & 48005 & 46903 & 44299 & 186093  \\
\hline
\end{tabular}
\end{table}

\subsection{Detector System}\label{subsec:detectors}
The NIR light collected by the JWST OTE and fed through the NIRSpec optical train is registered by two Hawaii-2RG (H2RG) sensor chip assemblies (SCAs), which are described in detail by \cite{rauscher07}. The SCAs are operated at a temperature of 42.8\,K, chosen as the best compromise between pixel operability and total noise of the detector system: a higher temperature leads to an increased number of hot pixels, while a lower temperature leads to higher noise in the signal chain.

As a reminder\footnote{for details, see the NIRSpec User Documentation at https://jwst-docs.stsci.edu/jwst-near-infrared-spectrograph/nirspec-instrumentation/nirspec-detectors}, the NIRSpec SCAs are non-destructively read “up-the-ramp”, using one of two fundamentally different readout modes: the so-called traditional readout mode (TRAD) or the improved reference sampling and subtraction \citep[IRS$^2$; pronounced ``IRS-square’’;][]{Rauscher17} readout mode. A number of detector subarrays with different sizes such as, for example, the ALLSLITS subarray are supported in TRAD mode only, offering faster readouts and using a slightly higher conversion gain to make the full physical well depth of the pixels accessible, which is of particular importance for time series observations of bright targets.

The in-orbit performance of the NIRSpec SCAs and their readout electronics is an important factor for the NIRSpec science performance, because the sensitivity of most NIRSpec observing modes is limited by the total detector noise in a given exposure \citep[see Appendix A in][]{jakobsen22}. The analysis of the detector performance data collected during the NIRSpec in-orbit commissioning campaign characterization has been presented in detail by \cite{birkmann22spie}. Here, we briefly summarize the most relevant results.

\subsubsection{Cosmetics} \label{subsec:cosmetics}
The vast majority of pixels in the NIRSpec detectors can be considered operable. Non-operable or bad pixels are, for example, those which have a very low quantum efficiency or no response at all, are not connected to the readout electronics, or exhibit a large / highly non-linear dark signal. The number of non-operable / bad pixels in the NIRSpec detectors as measured during commissioning is summarized in Table~\ref{tab:cosmetics}.

\begin{table*}
\caption{Summary of bad pixel statistics for the two NIRSpec detectors. Note that the total number of bad pixels can be less than the sum of the individual categories, as some bad pixels belong to several categories. Note that the fraction of operable pixels listed in the last row is for the active area of each detector (2040 x 2040 pixels).}\label{tab:cosmetics}
\centering
\renewcommand{\arraystretch}{1.2} 
\begin{tabular}{|l|r|r|l|}
\hline
\multicolumn{1}{|l|}{Bad pixel type} & \multicolumn{1}{c|}{NRS1} & \multicolumn{1}{c|}{NRS2} & \multicolumn{1}{c|}{Notes}\\
\hline
Open & 294 & 252 & Very low response, signal ends up in adjacent pixels (next row) \\
Adjacent open & 1664 & 1216 & Impacted by open neighbor pixel (previous row)\\
Dead & 7757 & 3938 & Does not respond to light \\
Low QE & 1766 & 887 & Low response to light \\
RC-like & 3908 & 1902 & Non-linear dark signal (RC-like ramp) \\
Hot & 6062 & 2123 & Large dark signal ($> 1$ e$^-\,s^{-1}$) \\
\hline
Total  & 16948 & 8275 & Total number of non-operable pixels \\
\hline
Operable pixels [\%]& 99.59 & 99.80 & Fraction of operable pixels \\
\hline
\end{tabular}
\end{table*}

\subsubsection{Dark Current}
The median dark signal of the two SCAs for the different NIRSpec readout modes is presented in Table\,\ref{tab:dark}. Two general observations can be made: i) the in-orbit dark signal for NRS1 is generally higher than for NRS2, which is fully in line with previous measurements obtained during the various ground testing campaigns.  ii) both NIRSpec SCAs exhibit a slightly elevated dark current signal compared to on-ground measurements, with a more pronounced increase towards the array edges \citep[see Fig.~2 in][]{birkmann22spie}. As discussed in \ref{sec:CR}, this increase is likely related to the cosmic ray environment at L2. However, even with this small increase compared to pre-launch measurements, the dark signal is still very low for most pixels, and not a driver for the total noise (and thus the sensitivity) of NIRSpec observations. Multiplexer readout glow is thought to dominate the observed dark signal \citep{regan2020}.

\begin{table}[htb]
    \caption{Median dark signal of the two NIRSpec SCAs (in e$^-$ per 1000s integration time) for traditional full frame, IRS$^2$ full frame, and ALLSLITS subarray readout modes, as measured during commissioning. For comparison, the equivalent numbers obtained during the last ground test campaign are listed in brackets.}\label{tab:dark}
    \centering
    \renewcommand{\arraystretch}{1.2} 
    \begin{tabular}{|l|c|c|c|}
    \hline
         &  \multicolumn{3}{c|}{Readout mode}\\ \hline
         SCA & TRAD & IRS$^2$ & ALLSLITS\\\hline 
         NSR1 & 9.0 (7.7) & 8.2 (7.1) & 22.3 (13.6)\\
         NRS2 & 6.9 (5.1) & 4.8 (4.0) & 15.3 (13.7)\\ \hline
    \end{tabular}
\end{table}

\subsubsection{Noise Performance}
The total noise for the different readout modes of the two NIRSpec SCAs as a function of effective integration time is summarized in Table\,\ref{tab:noise}. These numbers include the effects of cosmic rays, i.e.\ broken ramps due to cosmic ray hits and early saturation for some pixels.

\begin{table}[htb]
    \caption{Total noise of the two NIRSpec detectors for different readout modes and effective integration times as measured during commissioning.}\label{tab:noise}
    \centering
    \renewcommand{\arraystretch}{1.2} 
    \begin{tabular}{|l|c|c|c|}
    \hline
        & \multicolumn{3}{c|}{Effective Integration Time} \\ \hline
         Readout / SCA & $\sim$950\,s & $\sim$1700\,s & $\sim$3560\,s 
         \\\hline
         TRAD / NRS1 & 6.9 & 7.4 & N/A \\
         TRAD / NRS2 & 7.3 & 7.7 & N/A \\\hline
         IRS$^2$ / NRS1 & 5.9 & 6.6 & 8.5 \\
         IRS$^2$ / NRS2 & 7.2 & 7.6 & 9.2 \\\hline
         SUB / NRS1 & 7.0 & 7.8 & N/A \\
         SUB / NRS2 & 7.0 & 7.5 & N/A \\ \hline
    \end{tabular}
\end{table}

\begin{figure}[htb]
    \centering
    \includegraphics[width=0.99\columnwidth]{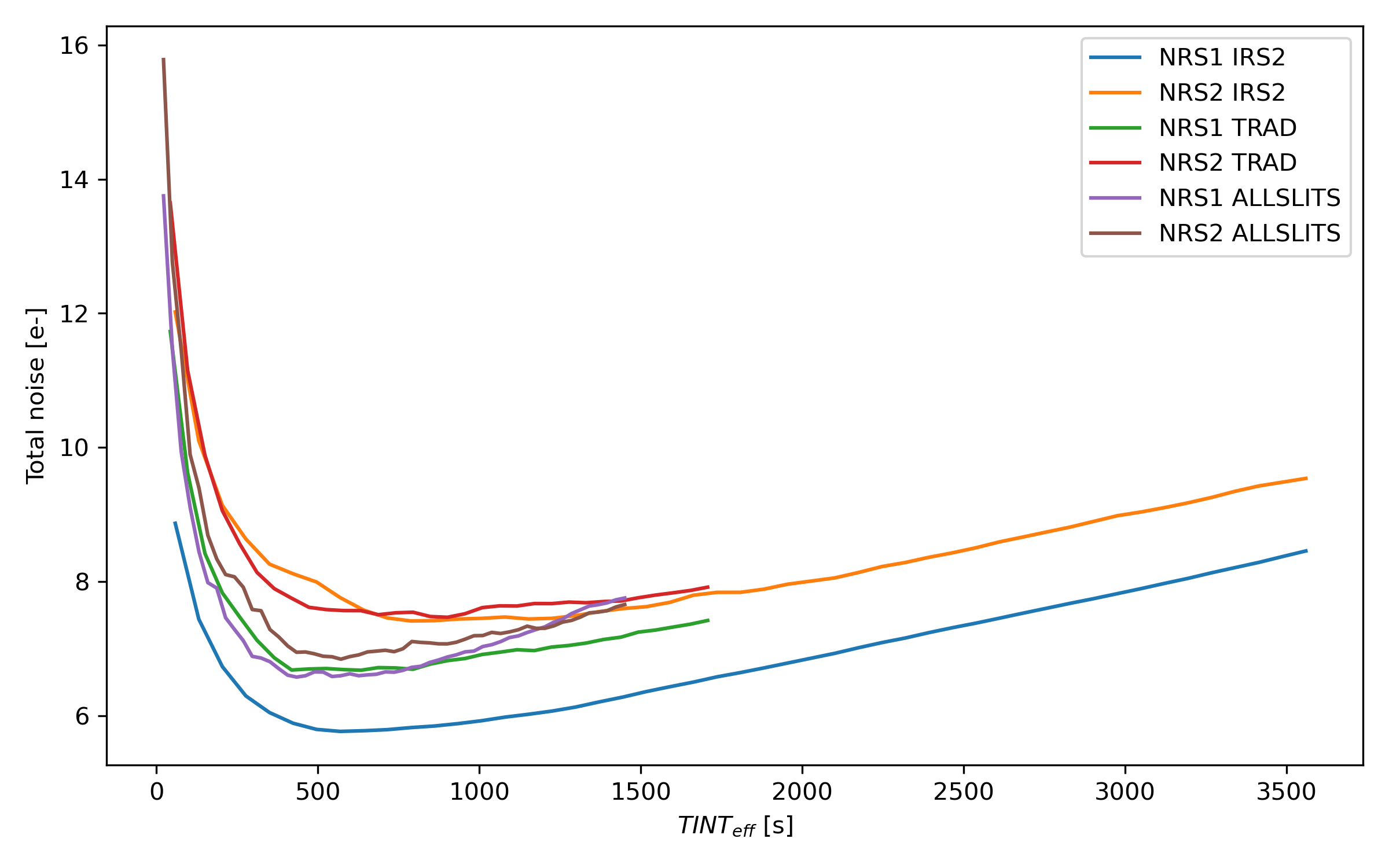}
\caption{Median total noise signal (in e$^-$) as a function of effective integration time for the two NIRSpec detectors and three readout modes: IRS$^2$, traditional full frame (TRAD), and ALLSLITS subarray.}
    \label{fig:detnoise}
\end{figure}

Figure\,\ref{fig:detnoise} shows the noise behavior as a function of integration time in more detail. For all readout modes, the total noise decreases with exposure length, up to integration times of $\approx$\,500\,s where it levels out and then slowly increases for longer integrations. Nevertheless, for observations of faint sources that are detector noise limited, it is still beneficial to use longer integration times for optimal signal-to-noise ratios. On the other hand, it is always advisable to have multiple integrations per observation, ideally in the form of dithered exposures to guard against early saturation after a strong cosmic ray hit, as well as to enable a robust rejection of outliers.

The IRS$^2$ readout mode shows the best total noise performance, in particular for NRS1, and should be used when observing faint targets. The ALLSLITS subarray readout mode has noise levels comparable to those in traditional full frame mode (TRAD), and is best used for observations of bright targets with short integrations, because the effective full well depth is higher due to the higher conversion gain, like for the other subarrays.

\subsubsection{Cosmic Ray Environment}\label{sec:CR}

From the NIRSpec dark exposures obtained during commissioning, we have measured an average CR hit rate of about $\rm 5.5\,cm^{-2}\,s^{-1}$, using the jump detection algorithm in the ramps-to-slopes pipeline. The observed hit rate is well in line with pre-launch predictions of the proton fluence at solar minimum \citep{giardino2019}. Most cosmic ray events affect several pixels (partially due to inter-pixel-capacitance), but are compact, with a typical hit area of about 10.5 pixels. However, there can also be longer streaks and so-called `snowballs', affecting many pixels at once. As described in \cite{birkmann22spie}, snowballs have the following characteristics:

\begin{itemize}
    \item a core of fully saturated pixels
    \item an extended `halo' around the saturated core that is detected at the same time (i.e. within the same correlated double sample), with the intensity of the halo dropping off towards larger radii
    \item often accompanied by a shower of more compact cosmic ray events in the vicinity
    \item core and halo are often spherical in appearance, but can be very elongated as well
\end{itemize}

The flux of snowballs varies, but stronger ones can produce tens of millions of electrons within their saturated core, indicating large energies of the involved particle(s). Snowballs are observed by all NIR instruments on JWST \citep[see e.g.][]{mrieke22}, but their origin is not yet fully understood.

\subsubsection{Persistence}
Image persistence, or latency signal, is an unavoidable effect in NIR HgCdTe detectors. It manifests itself as an 'afterglow' in pixels that have been subjected to strong illumination, resulting in a faint residual image in the following integrations. Persistence is caused by electrically active defects, so-called 'charge traps', in the detector material which accumulate charge that is only released in later exposures. Past results from the ground testing of the JWST NIR detectors has shown that their persistence behavior is generally better than that of similar detectors on previous NASA missions including the Hubble Space Telescope: any persistence decays to below the background level after about 2000\,s \citep{Rauscher14}. While the detailed characterization of the persistence behavior of the two NIRSpec SCAs is still ongoing, the in-orbit data collected so far suggest that persistence is not a major concern for most NIRSpec science programs.

\section{NIRSpec Science Performance} 
Many key parameters of the end-to-end science performance of NIRSpec and the JWST OTE could only be measured with on-sky data obtained after launch. In this section, we discuss a number of aspects that are relevant for the scientific performance of all NIRSpec modes.

\subsection{Photon Conversion Efficiency and Sensitivity}\label{subsec:pce}
For NIRSpec, and in fact any optical instrument, a critical performance parameter is its efficiency, i.e. the fraction of photons incident on the primary mirror that are being registered by the detector after passing through the entire optical path. This metric, together with the noise performance of the NIRSpec detectors, drives the ultimate sensitivity for astronomical observations.

\begin{table}[htb]
   \caption{NIRSpec spectral configurations.}\label{tab:configs}
   \centering     
	\begin{tabular}{|c|c|c|c|c|}
	\hline
	Band & Disperser & $\lambda / \delta\lambda$ & Filter & $\lambda$ range [$\mum$] \\
	\hline
       0 & $\begin{tabular}{@{}c@{}} G140M \\ G140H \end{tabular}$ & $\begin{tabular}{@{}c@{}} 1000 \\ 2700 \end{tabular}$ & F070LP & 0.7 -- 1.2\tablenote{the Band 0 configurations using the F070LP filter will obtain spectra over a much wider wavelength range which, however, contain 2nd-order contamination beyond $1.2\mum$. Users who are prepared to deal with this contamination can potentially use Band 0 out to $\approx 1.8\mum$.} \\
       \hline
	   I & $\begin{tabular}{@{}c@{}} G140M \\ G140H \end{tabular}$ & $\begin{tabular}{@{}c@{}} 1000 \\ 2700\end{tabular}$ & F100LP & 1.0 -- 1.8 \\
        \hline
	   II & $\begin{tabular}{@{}c@{}}G235M \\ G235H \end{tabular}$ & $\begin{tabular}{@{}c@{}} 1000 \\ 2700\end{tabular}$ & F170LP & 1.7 -- 3.1 \\
        \hline
	   III & $\begin{tabular}{@{}c@{}}G395M \\ G395H \end{tabular}$ & $\begin{tabular}{@{}c@{}} 1000 \\ 2700\end{tabular}$ & F290LP & 2.9 -- 5.2 \\
        \hline
	   n/a & PRISM & 30-330 & CLEAR & 0.6 -- 5.3 \\
	\hline
	\end{tabular}
\end{table}

Because NIRSpec is a complex instrument that supports many different observing modes, its optical train is rather complicated, as evident from Fig.~\ref{fig:layout}. Except for the order-separation filters and the low resolution double-pass prism, the NIRSpec optics are reflective throughout. Photons captured by the JWST primary mirror undergo a total of 19 reflections before reaching the NIRSpec detector array in the MOS, FS, and BOTS observing modes. For IFS mode, there are an additional 8 reflections in the IFU optics \citep[see][]{boeker22}. Cleanliness and high reflectivity of the NIRSpec optics therefore was of ultimate importance throughout the construction and test phases.

By necessity, the metric of choice to measure the optical throughput is the Photon Conversion Efficiency (PCE), i.e. the ratio of photons incident on the JWST primary mirror to electrons registered by the NIRSpec detector system. It can be derived using observations of a `flux standard' (typically a well-characterized star) with an accurately known spectral energy distribution (SED). As can be seen from Table\,\ref{tab:configs}, NIRSpec has a total of nine spectral configurations, and the throughput must be measured separately for each of them.

As described in more detail by \cite{giardino22}, the NIRSpec optical efficiency meets or exceeds the pre-flight expectations. In particular, significantly higher than predicted
PCEs are achieved for the high-resolution
configurations, for the MOS/FS mode, and for all configurations below
$\sim2.6~\mum$ for the IFS mode. The 10-20\% lower than expected efficiencies above $\sim 4~\mum$, apparent in particular for the IFS
mode, are (mostly) explained by the more significant path losses at longer wavelengths, which are reflected in the measurements but not the predictions \citep[see Fig. 3 in][]{giardino22}.

These in-orbit PCE measurements , together with the detector noise performance discussed in Section\,\ref{subsec:detectors}, can be used to calculate the in-flight NIRSpec sensitivity for its various science modes and configuration, following the methodology described in Appendix A of \cite{jakobsen22}. Figure\,\ref{fig:sens} shows the results of the calculations in terms of continuum sensitivity curves for a point source observed in the MOS and IFS modes, using a bench-mark observation in all spectral configurations. The achieved level of sensitivity is extremely impressive when compared to other NIR spectrographs with similar observing modes and spectral resolutions: by at least two orders of magnitude, NIRSpec is the most sensitive NIR spectrograph currently available for astronomical studies \citep[see also][]{rigby22a}.

\begin{figure*}[htb]
    \centering
    \includegraphics[width=0.99\columnwidth]{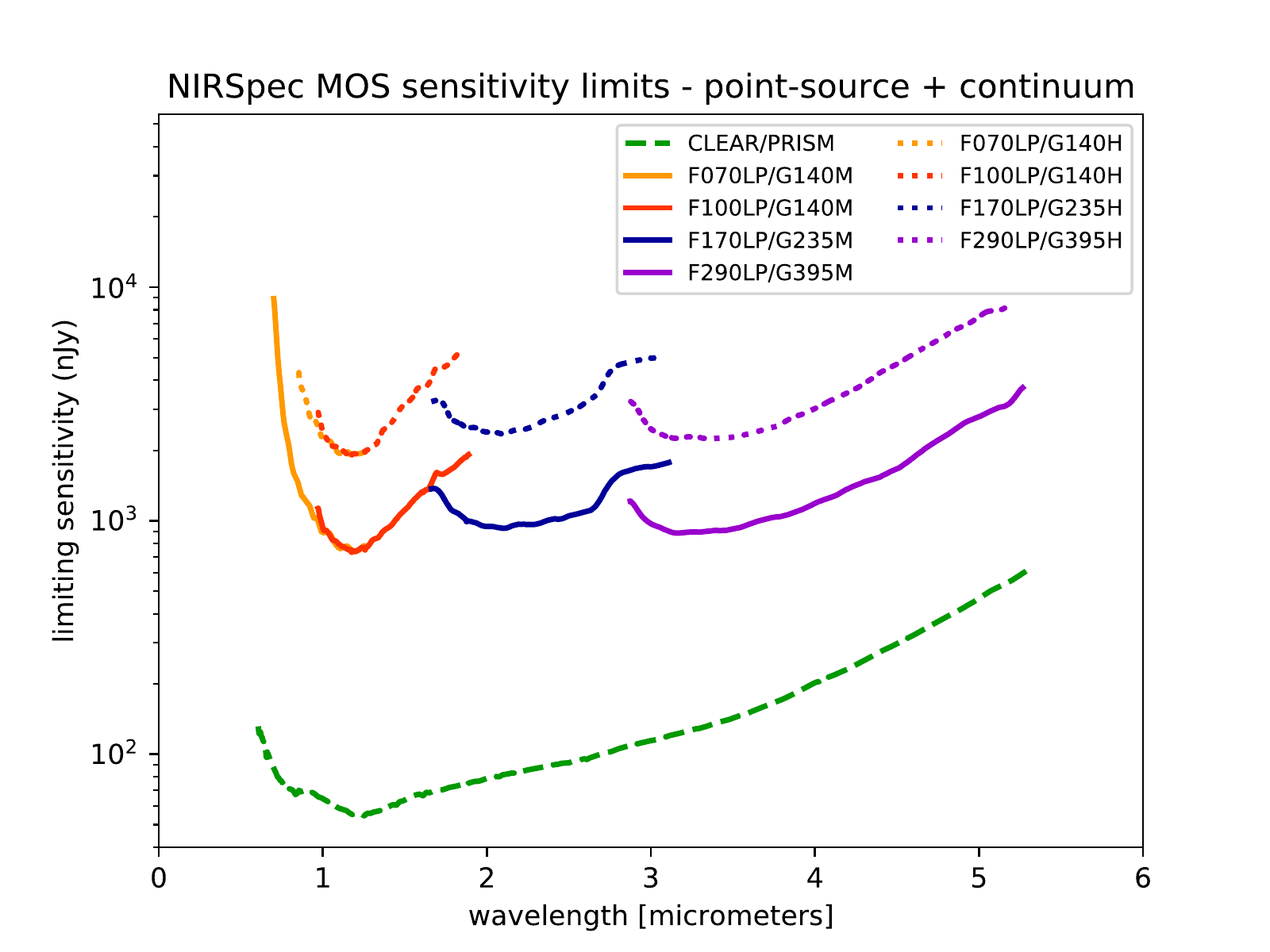}
    \includegraphics[width=0.99\columnwidth]{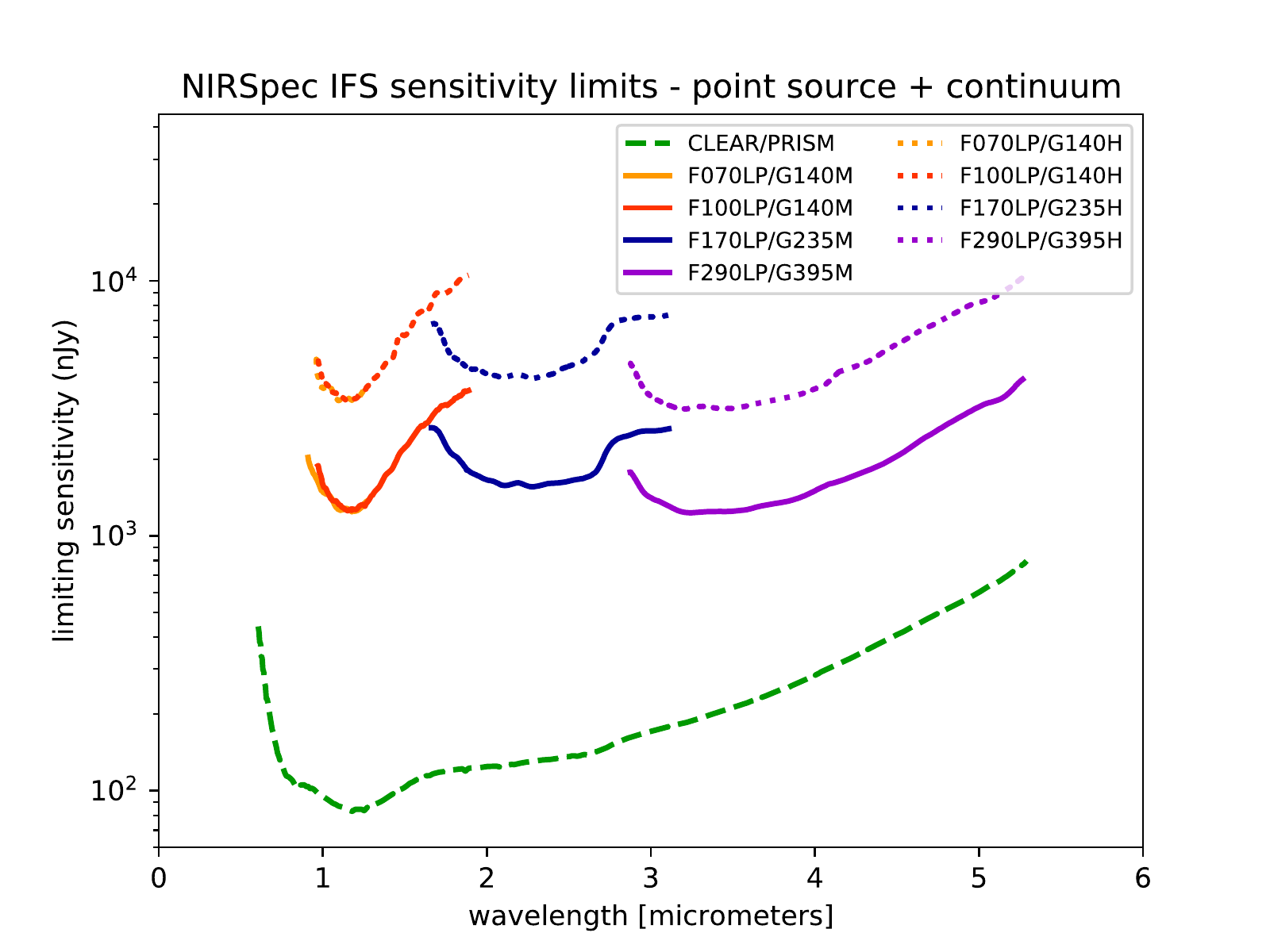} \\
\caption{NIRSpec point-source continuum sensitivity in MOS/FS (left) and IFS (right) mode, derived from in-orbit measurements and assuming a well-centered source in a microshutter or an IFU slice, respectively. The sensitivity for the S200 slits in FS mode is similar to the one in MOS mode. The plots show, for each disperser and as a function of wavelength, the flux required to reach $\rm S/N=10$ per spectral pixel in 10,000s. More specifically, the computations assume 10 NRSIRS2RAPID exposures of 1006.7~s each (70 groups of 1 frame), using the methodology described in Appendix~A of \cite{jakobsen22}.}
    \label{fig:sens}
\end{figure*}

\subsection{Spectro-Photometric Calibration}
The full photometric calibration of NIRSpec requires three major steps in the reduction pipeline, with their associated reference files: i) the detector flat (D-Flat) to capture the pixel-to-pixel response variations, and derived from component-level ground test data of the two NIRSpec SCAs, ii) the spectrograph flat (S-Flat) to correct the throughput variations of the spectrograph optics, and measured from exposures using the internal calibration lamps in the CAA, and iii) the FORE optics flat (F-Flat) to characterize any field- and wavelength-dependent effects caused by the OTE and the NIRSpec FORE Optics. For the last step, observations of spectro-photometric standard stars are necessary to verify the integrity of the entire NIRSpec optical system, in particular the pick-off and FORE optics (including the FWA), neither of which can be illuminated with the CAA lamps \citep[see][for a detailed description of the light path from the CAA]{tePlate05}. For a detailed overview of the NIRSpec flat field strategy we refer to \citet{rawle16}.

\begin{figure*}[htb]
    \centering
    \includegraphics[width=0.95\textwidth]{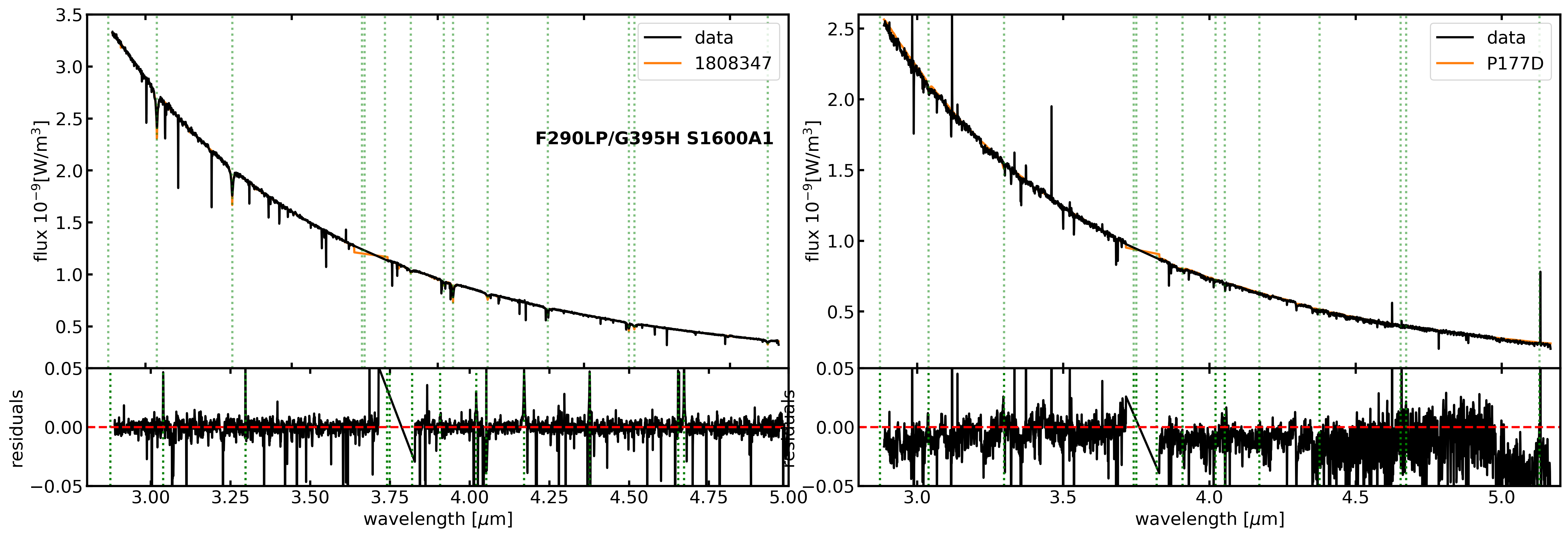}
\caption{The spectra of standard stars 1808347 (left) and P177D (right), extracted with the NIRSpec-internal reduction pipeline, and compared to the CALSPEC model to verify the flux calibration. Both spectra were taken through the S1600A1 slit using the F290LP/G395H configuration. The bottom frames show the residuals. The observations of P177D were obtained without a target acquisition step, and the systematically lower flux compared to the model is most likely due to imperfect centering in the S1600A1 aperture, which can easily cause aperture losses of 1-2\%.}
    \label{fig:flux_calib}
\end{figure*}

The S-Flat step corrects any throughput variations in the spectrograph, i.e. introduced in the optical path after the FWA and before the detector. The S-Flat reference files for all NIRSpec observing modes are derived from internal lamp exposures and after correction for the detector response (D-Flat) and spectral energy distribution of the lamp(s). Because the throughput is strongly dependent on the properties of the chosen disperser and slit aperture, different reference files are created per detector for each combination of prism or grating and NIRSpec observing mode (FS, IFS, or MOS). For the FS and IFS modes, the S-flat consists of two parts: (i) a 2D image capturing the pixel-to-pixel variations for all slits/slices, and (ii) a vector for each slit and the IFU, to correct the fast throughput variations with wavelength. This approach is possible for the FS and IFS mode, because for a given aperture and spectral configuration, each detector pixel is always illuminated by the same wavelength (modulo the non-repeatability of the GWA, which is corrected separately using the position sensors described in Sec.\,\ref{subsec:wheels}).

For MOS mode, in contrast, the same detector pixel can receive light of different wavelengths, depending on the position of the open MSA shutter. Hence, the MOS S-flat reference consists of a data cube, sampling the wavelength variation for each pixel, so that the specific S-flat for any given MSA configuration can be derived on-the-fly by the pipeline. Given the large number of shutters and the finite amount of time available for commissioning, direct measurement of the throughput for every shutter was impossible. Therefore, only a subset of MSA shutters was observed, from which the complete (smoothed) S-flat cube could be generated by interpolation.

Similar to the S-Flat, the F-Flat must be created for each of the filter and grating wheel combinations listed in Table\,\ref{tab:modes} and for each of the three observing modes (FS, IFS, and MOS\footnote{The in-flight update for the MOS mode is still pending, but the necessary data have been taken.}), to take into account any wavelength dependence on the throughput of the OTE and FORE optics. Standard star observations were obtained for all but the F100LP/G140M configuration, for which calibration files based on simulated data will remain in place until such observations are taken during Cycle 1.  

We used the A3V star 1808347 (2MASS J18083474+6927286) for all gratings, while for the CLEAR/PRISM configuration, we used either the A8III star 1743045 (2MASS J17430448+6655015), or the G0-5 star SNAP-2 (2MASS J16194609+5534178), because 1808347 would saturate the low-resolution spectra. For the fixed slits, we used a 3-nod pattern (2-nod pattern for S1600A1) to obtain the data, and a 4-nod pattern with a 4 spaxel (0.4\,arcsec) extraction radius for the IFS observations. The MOS program was executed using a 3-shutter nod pattern for each of the filter/grating combinations.

To derive the conversion to absolute flux units (which is part of the F-Flat step), the extracted spectra were divided by the resampled standard star templates available on the CALSPEC website\footnote{\url{https://www.stsci.edu/hst/instrumentation/reference-data-for-calibration-and-tools/astronomical-catalogs/calspec}} \citep{bohlin2014}. To create a smooth F-Flat, any outliers such as remaining hot pixels, stellar absorption lines, or the 0$^{\rm th}$ order contamination of the lamp spectra used to create the S-Flat, were masked and a smooth vector was fitted, creating the final F-Flat.

To verify the validity of the NIRSpec calibration approach, we re-extracted the spectrum of the standard stars using the NIRSpec-internal data reduction pipeline, and compared the resulting spectra to the CALSPEC template, as shown in Fig.\,\ref{fig:flux_calib} for the case of FS mode with the S1600A1 aperture.  Overall, the agreement is very good, with an RMS of the residuals well below 2\%. For the other NIRSpec modes and apertures,  the results are of similar quality. Note that this comparison only represents an estimate of the `best case' calibration accuracy, because it is performed on the same star that was used to derive the pipeline reference files, and thus does not account for any systematic uncertainties. 

While an evaluation of the systematic errors must wait for additional calibration data obtained during Cycle 1 or later, commissioning spectra of two other standard stars, WD1057+719 (a DA1.2 white dwarf, M-gratings) and P177D (a G0V star, H-gratings), were obtained in FS mode with the S1600A1 slit (see Table\,\ref{tab:modes}). These observations were executed early in commissioning and before the wide aperture TA procedure (see Section\,\ref{subsec:WATA}) was available. Therefore, a proper centering is not guaranteed and (small) uncorrected aperture losses are possible. Nevertheless, we measure residuals between the flux template and the extracted spectra of $-2.12\pm2.55\%$, $-3.76\pm1.00\%$, $-3.67\pm1.57\%$, $-5.09\pm1.16\%$, $-1.00\pm1.21\%$, and $-0.91\pm1.97\%$ for the F070LP/G140H, F170LP/G235M, F170LP/G235H, F290LP/G395M , F290LP/G395H, and CLEAR/PRISM configurations, respectively. 

These residuals are well below the pre-launch requirement ($10\%$ absolute photometric accuracy of all NIRSpec spectra, i.e. even after correction for `delta' slit losses due to imperfect source centering). While source centering is most critical for the 200\,mas wide slits, it is worth mentioning here that after a successful target acquisition procedure (see Section\,\ref{sec:TA}), the source is typically placed within 10\,mas of the slit center. This magnitude of source displacement would result in insignificant `delta' slit losses.

As noted above, the results presented here were derived using the NIRSpec-internal data reduction pipeline. The user pipeline provided by STScI is currently being checked and improved to deliver similar results, and the status is discussed further in Section\,\ref{sec:pipeline}.

\begin{figure*}[htb]
    \centering
    \includegraphics[width=0.95\textwidth]{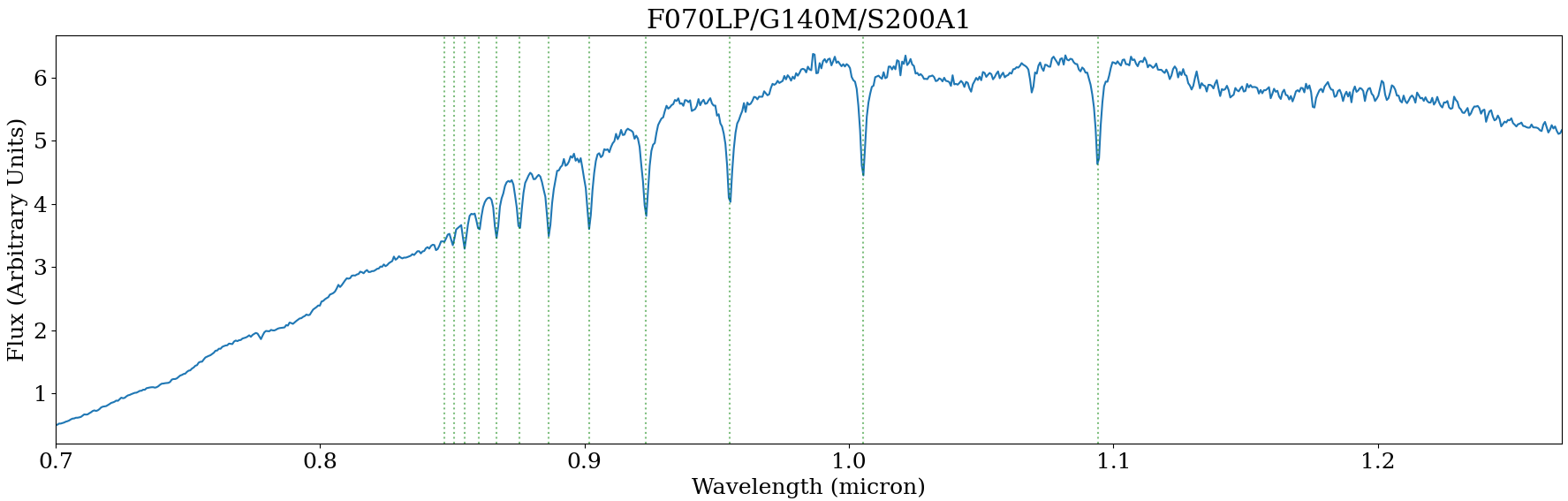} \\
    \includegraphics[width=0.95\textwidth]{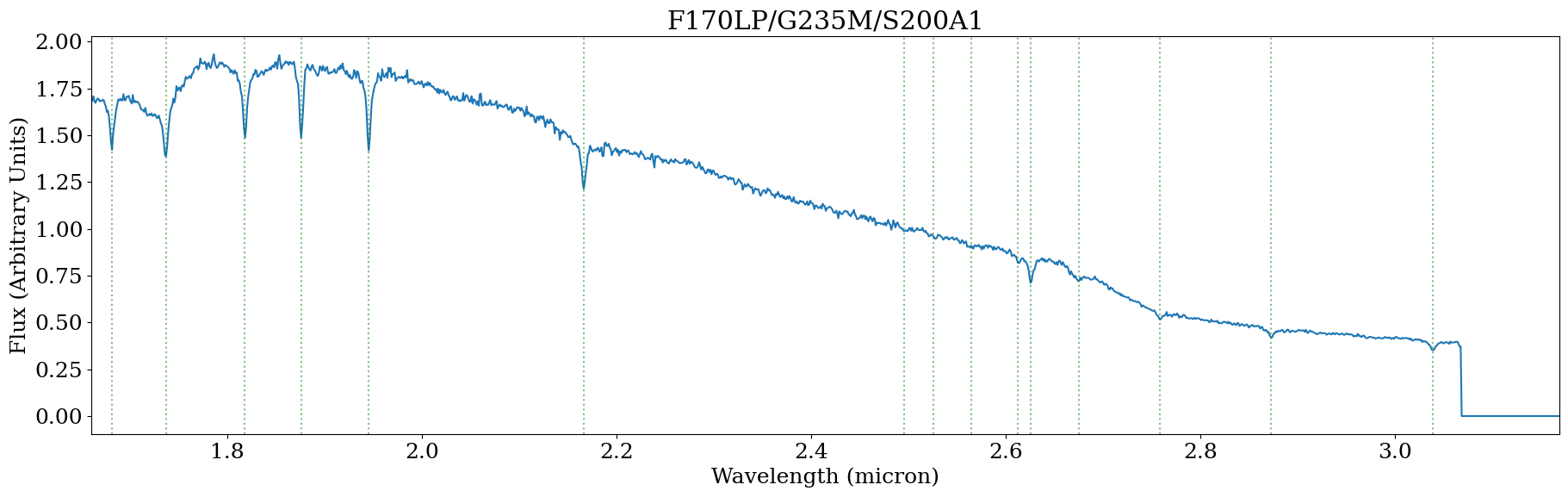} \\
    \includegraphics[width=0.95\textwidth]{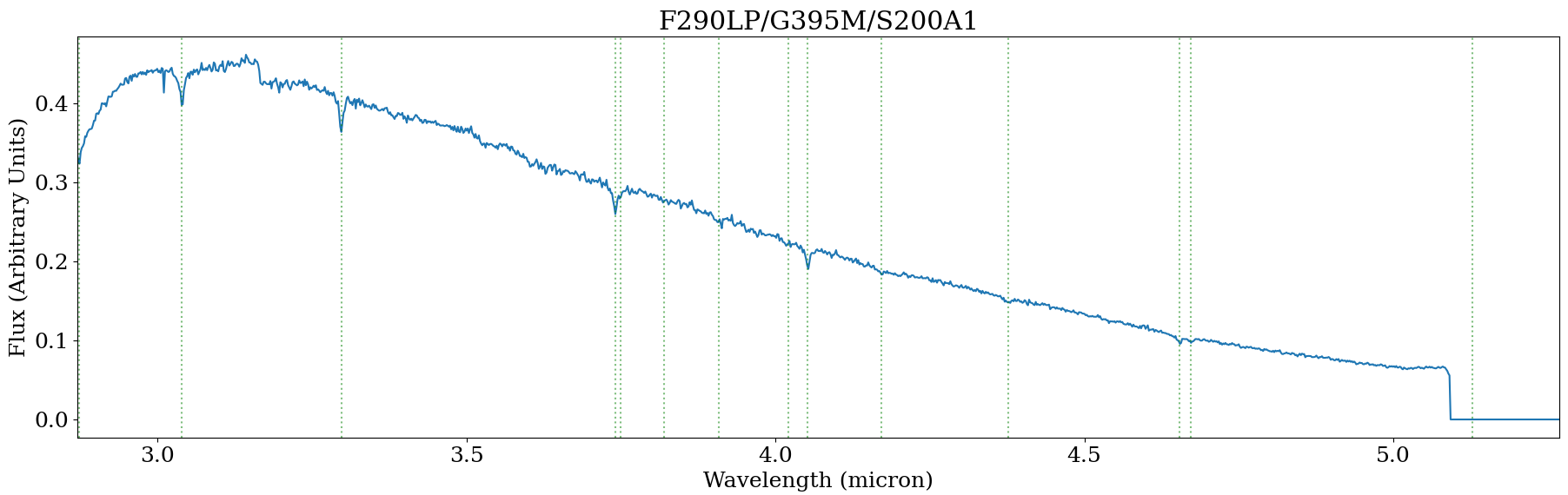} 
\caption{Wavelength-calibrated spectra of the spectrophotometric standard star 1808347, obtained in FS mode through the S200A1 slit (see Table\,\ref{tab:modes}), and using the medium-resolution gratings in three of the four spectral configurations (F100LP/G140M is not shown). These data were obtained during the NIRSpec commissioning program `Spectrophotometric Sensitivity and Absolute Flux Calibration' (PID1128). }
    \label{fig:wavecal}
\end{figure*}

\subsection{Wavelength Calibration}\label{subsec:wavecal}
The accuracy of the NIRSpec wavelength calibration is determined by the performance of the parametric instrument model, which was a crucial product to be derived from NIRSpec commissioning data. As described in \cite{luetzgendorf22}, the residuals in the final instrument model were well within the requirements for the wavelength and astrometric calibration of NIRSpec data. This is illustrated in Fig.~\ref{fig:wavecal}, which shows wavelength-calibrated spectra of the spectrophotometric standard 1808347, an A3V star with prominent hydrogen absorption features, taken through the 200\,mas wide fixed slits.

As can be seen, the NIRSpec instrument model allows for a highly accurate wavelength calibration, with residuals for internal lamp spectra well below the requirement (1/8 of a resolution element) for all gratings analyzed so far. Note that the full verification for all pipeline products in the various NIRSpec modes is still ongoing, especially in the case of the IFS mode. 

\subsection{Photometric Stability for Time-Series Observations}\label{subsec:TSO}
The temporal stability of the instrument response to a constant flux stimulus is a critical parameter for the accuracy of measured light curves of astronomical targets, e.g. in the case of Time Series Observations (TSOs) of transiting exoplanets. In order to quantify the stability of the NIRSpec response over various timescales, we obtained a TSO of the star HAT-P-14 (PID 1118; PI: Proffitt) during the transit of its well-characterized exoplanet HAT-P-14\,b using the G395H/F290LP grating/filter combination (see Table\,\ref{tab:configs}). The observations, which also served the purpose of verifying the correct execution of the BOTS observing template \citep[see][]{birkmann22a}, were analyzed in detail by \cite{espinoza22}.

To recap, the transmission spectrum of HAT-P-14\,b could be measured with a precision of 
50-60 ppm at $R=100$ (rebinned down from $R=2700$), which is in excellent agreement with pre-flight expectations, and close to the photon-noise limit for a $J = 9.094$, F-type star like HAT-P-14. There were two noteworthy features observed in this analysis. The first was a weak linear trend in the white-light response as a function of time, which was observed to be stronger in 
NRS1 (-150 ppm/hour) than in NRS2 (30 ppm/hour). This was shown to also 
be slightly wavelength dependant in NRS1, but not in NRS2. These trends can be easily corrected for, and are most likely caused by low-amplitude instabilities in the detector signal chain. The second is the fact that 
binning pixels in the spectral direction seems to not decrease the noise 
level in the time-series as $1/\sqrt{N_{\textnormal{bin}}}$ where $N_{\textnormal{bin}}$ is the size of the bin. It is likely this is related to unaccounted covariance between pixels due to effects such as, e.g., 1/f noise. This can also be accounted for by either 
simply working with the added pixels on each spectral channel or 
considering a more complex spectral extraction and binning scheme 
that includes this spatial covariance of pixels in the detector \citep[see, e.g.,][]{schlawin}.

Given the excellent performance of the BOTS mode, and the absence of any strong variations and systematic trends in the combined response from telescope and instrument over timescales of a few hours to a day, NIRSpec promises to fulfill its key role for cutting-edge transiting exoplanet atmospheric science with JWST during Cycle 1 and beyond.

\begin{figure*}[htb]
    \centering
    \includegraphics[width=0.96\textwidth]{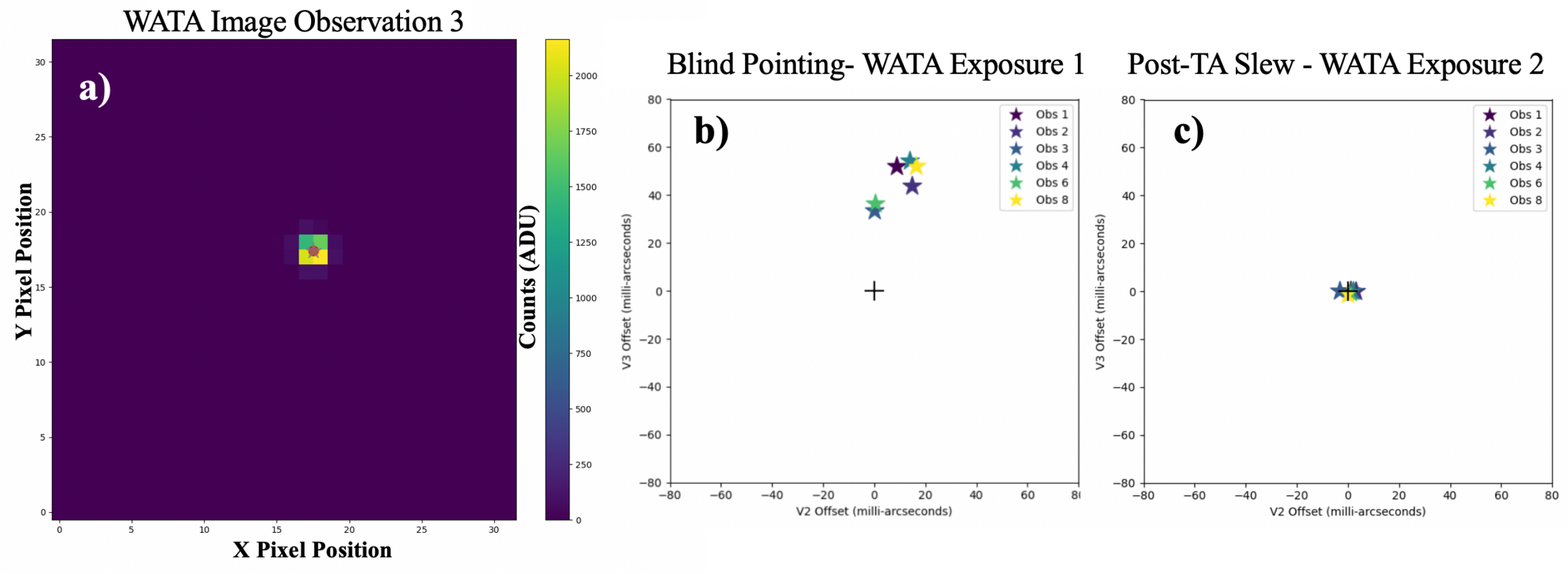}
\caption{(a) The Wide Aperture Target Acquisition (WATA) exposure image (from PID 1118 Obs 3).  The 32 pixel by 32 pixel aperture is placed around the 1.$"$6 $\times$ 1.$"$6 (approximately 16 by 16 pixels) square aperture, but not exactly centered.  (b) The V2, V3 coordinate offset of the initial blind pointing position of the star that is calculated for the first exposure for each WATA Observation test.  The average blind pointing for the star is better than 50 mas radial from the wide aperture center. (c) The post-target acquisition pointing showing the improved centering of the TA target.  The average corrected position for the stars is better than 2.5 mas radial. }
    \label{fig:wata}
\end{figure*}

\subsection{Target Acquisition}\label{sec:TA}
Depending on their scientific needs, NIRSpec users have a number of options for fine-tuning of the telescope pointing after the completion of the initial slew to the target field, which typically places the source within 100\,mas of its intended position. Programs for which this level of accuracy is sufficient do not need a dedicated TA procedure, and should select TA=NONE or TA=VERIFY\_ONLY. The difference between these two options is that (for MOS and IFS mode) the latter obtains an undispersed image of the sky after the spectroscopic exposures in order to allow a determination of the pointing post-facto.

Programs that require a more accurate target placement
of an individual source in either the IFU, one of the fixed slit apertures, or the MOS single-point field position should use the NIRSpec Wide Aperture Target Acquisition (WATA) procedure. For MOS observations, on the other hand, the added complexity of requiring a roll angle optimization calls for the use of a dedicated method called Micro-Shutter Array Target Acquisition (MSATA). The details and in-orbit performance of both of these TA methods are discussed in the following subsections.

\subsubsection{Wide Aperture Target Acquisition}\label{subsec:WATA}
 The WATA procedure takes an image of an isolated, point-like target through the 1.$"$6 $\times$ 1.$"$6 wide fixed slit aperture (S1600A1). Using this image, the onboard software then computes the centroid of the source emission to determine its position after the initial `blind' telescope slew, and autonomously calculates the corrective `delta' slew required to accurately position either this target or another nearby target at the optimal location in the NIRSpec science aperture. The total duration of the WATA procedure can be as short as 5\,min, and as long as 11\,min, depending on the size of the detector area being used (subarray or FULL frame), and the depth of the acquisition exposure.

Figure\,\ref{fig:wata}a shows the WATA image of a star used to verify the WATA process during commissioning, obtained from Proposal ID 1118 (PID1118). This program contained six successful WATA attempts\footnote{The data discussed here are from Observations 1-4, 6, and 8.} that were used to evaluate the onboard algorithms, the correct computation and execution of the offset slew, and the subsequent science exposures. Figure\,\ref{fig:wata}b shows the position of the source after the initial slew, i.e. at the beginning of each WATA. Because of the excellent slew and pointing performance of the observatory, the WATA starting position is already quite close to the aperture center: on average, the target is within 50\,mas radial distance, or one half of the size of a NIRSpec detector pixel. 

Figure\,\ref{fig:wata}c shows the target position at the end of the WATA procedure, i.e. after the corrective slew to the center of the S1600A1 aperture was computed by the onboard algorithm, and executed by the telescope guiding system. For the six successful WATA observations in PID1118, the average target position was within 2.5\,mas from the aperture center, which is almost ten times better than the pre-launch requirement of 20\,mas.  

For observations that require the science target to be placed in apertures other than the S1600 slit itself, further tests executed during commissioning and the early science program have demonstrated that the telescope slew accuracy is sufficient to place the target at, e.g., the intended location in the IFU aperture with an accuracy of better than 10 mas, again well within the requirement.  

\subsubsection{Micro-Shutter Array Target Acquisition}\label{subsec:MSATA}
The NIRSpec MOS mode requires that the images of astrophysical targets are accurately placed onto the MSA over the entire ($3\arcmin \times 3\arcmin$) FOV, such that they fall within their dedicated microshutters. The onboard algorithms to achieve this rely on highly precise coordinate transformations between the detector, MSA, and sky planes, which are another crucial output of the NIRSpec parametric instrument model \citep{luetzgendorf22}. 

As explained in detail by \cite{keyes18}, the NIRSpec MSA target acquisition (MSATA) process uses a set of 5 to 8 `reference stars' that are imaged onto the detector through the microshutter grid. For MSATA exposures, the MSA can be either in the all-open configuration or in `protected' mode, which closes the shutters around bright stars to prevent persistence in the subsequent science exposures. Two MSATA exposures are acquired, separated by an offset equivalent to half of the microshutter pitch (in both x and y direction), in order to mitigate the effect of vignetting by the MSA bars. The centroid position of each reference star is calculated autonomously by the on-board software, and the set of centroids is analyzed, outliers are clipped, and their mean offset from the intended position is used to correct the initial spacecraft pointing (`pitch' and `yaw') and position angle (`roll').  

The duration of the MSATA procedure depends on the number of reference stars used and the depth of the acquisition exposures, ranging from 23\,min for the minimum number of reference stars (5) and the shortest readout pattern, to 35\,min for 8 reference stars (the recommended number), and the deepest available readout pattern. In cases where the MSATA algorithm does not converge to a sufficiently accurate solution, it may be repeated once, adding up to 22\,min to the duration if the maximum of 8 reference stars is used.

\begin{figure*}[htb]
    \centering
    \includegraphics[width=0.98\textwidth]{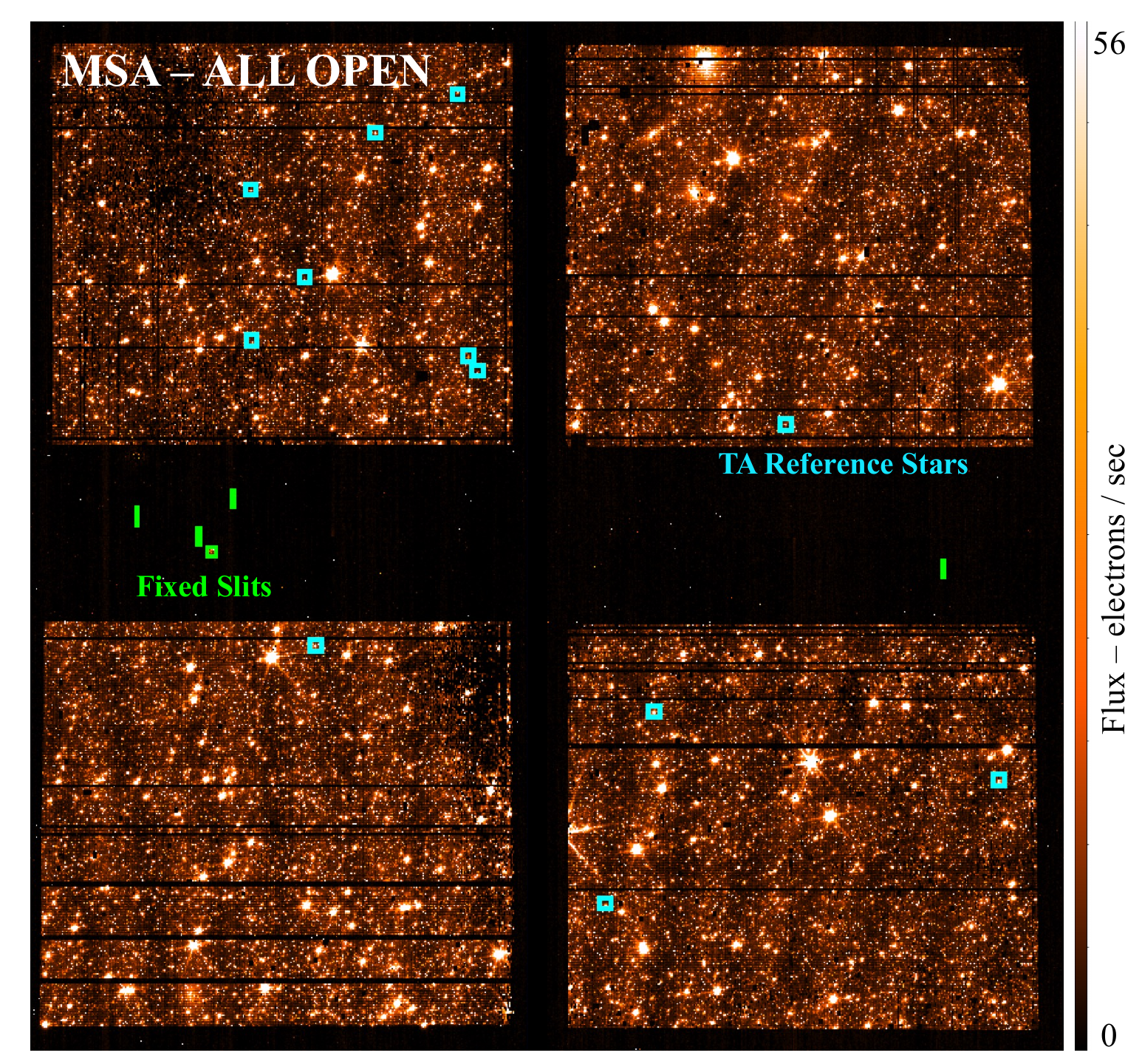}
\caption{Undispersed NIRSpec exposure of the JWST astrometric field (from PID 1117, Obs 31).  This crowded, but astrometrically well calibrated stellar field was selected to perform the first in-orbit test of the MSATA procedure.  To ensure that enough stellar centroids were successfully measured, this observation used a special requirement to allow the use of 12 TA reference stars (highlighted by the cyan boxes), instead of the default maximum of 8 stars. The NIRSpec Fixed Slit apertures are highlighted in green.}
    \label{fig:msata}
\end{figure*}

The NIRSpec MSATA process is the only operational situation where the roll angle of the Webb telescope is adjusted after the start of an observation. To provide an impression of the complexities involved, Figure\,\ref{fig:msata} shows the image used to verify the NIRSpec MSATA process during commissioning (PID 1117, Obs. 31). The target field is the astrometric calibration field \citep{anderson21} in the Large Magellanic Cloud (LMC) which had been pre-observed with HST and ground-based telescopes to provide accurate stellar coordinates for the astrometric calibration of the JWST focal plane. 
In such highly crowded fields, careful planning is necessary to identify the optimal MSATA reference stars, which must not saturate in the TA exposure, and must be suitably isolated for optimal centroid calculation by the on-board algorithm. 

To allow accurate calibration of MOS spectra, the MSATA process must place the science targets with an accuracy of just 20\,mas across the NIRSpec FOV. The most important driver for the MSATA accuracy is the absolute anchoring of the planning catalog in rotation, as it is critical for the proper derivation of the TA roll solution.  If the desired accuracy of the target placement in the shutters can be relaxed, the MSATA can also be planned using catalogs with poorer astrometric precision or rotation anchoring, or using galaxies with compact central cores instead of stars.  

\begin{figure*}[htb]
    \centering
    \includegraphics[width=0.98\columnwidth]{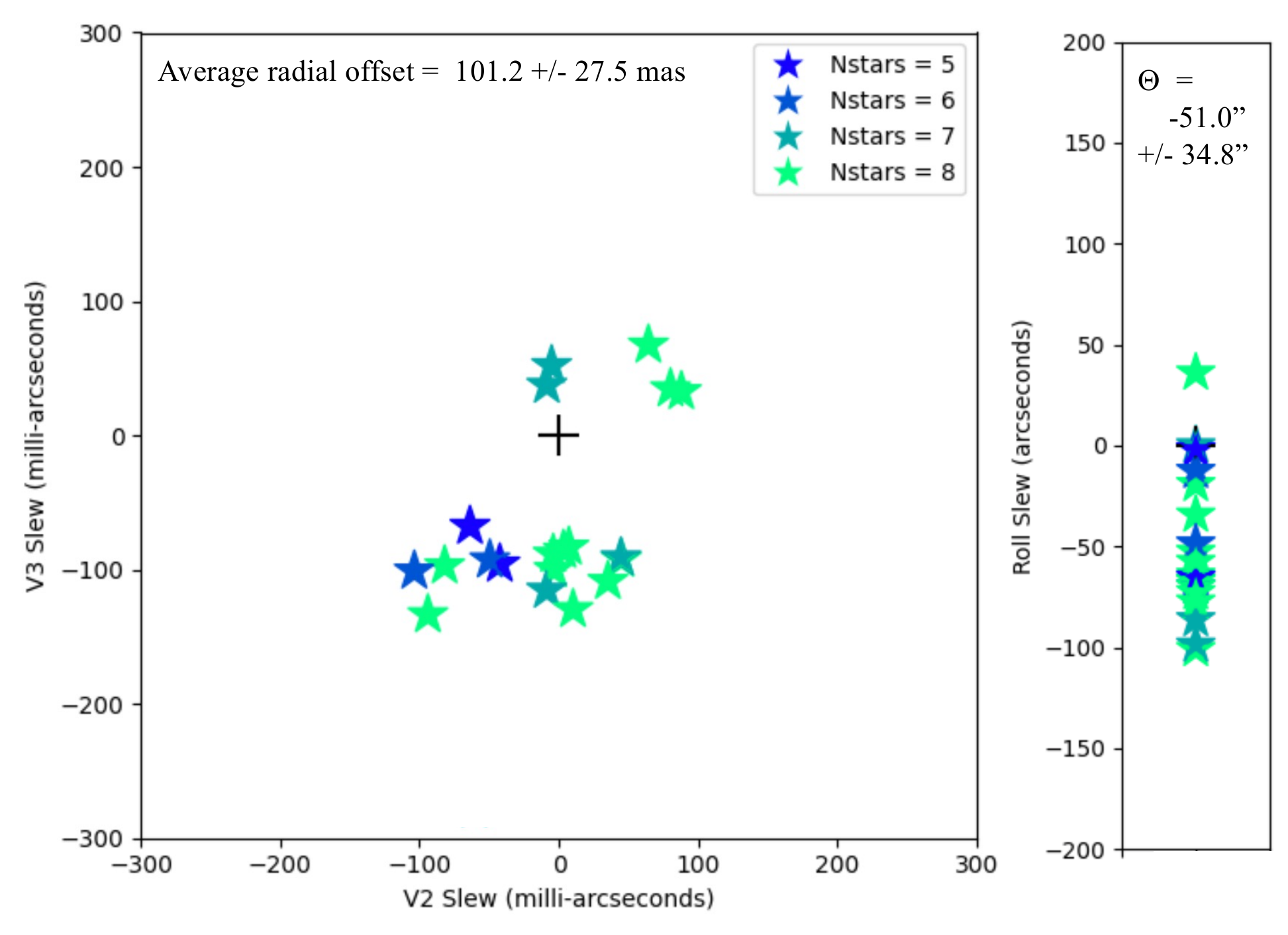}
    \includegraphics[width=0.98\columnwidth]{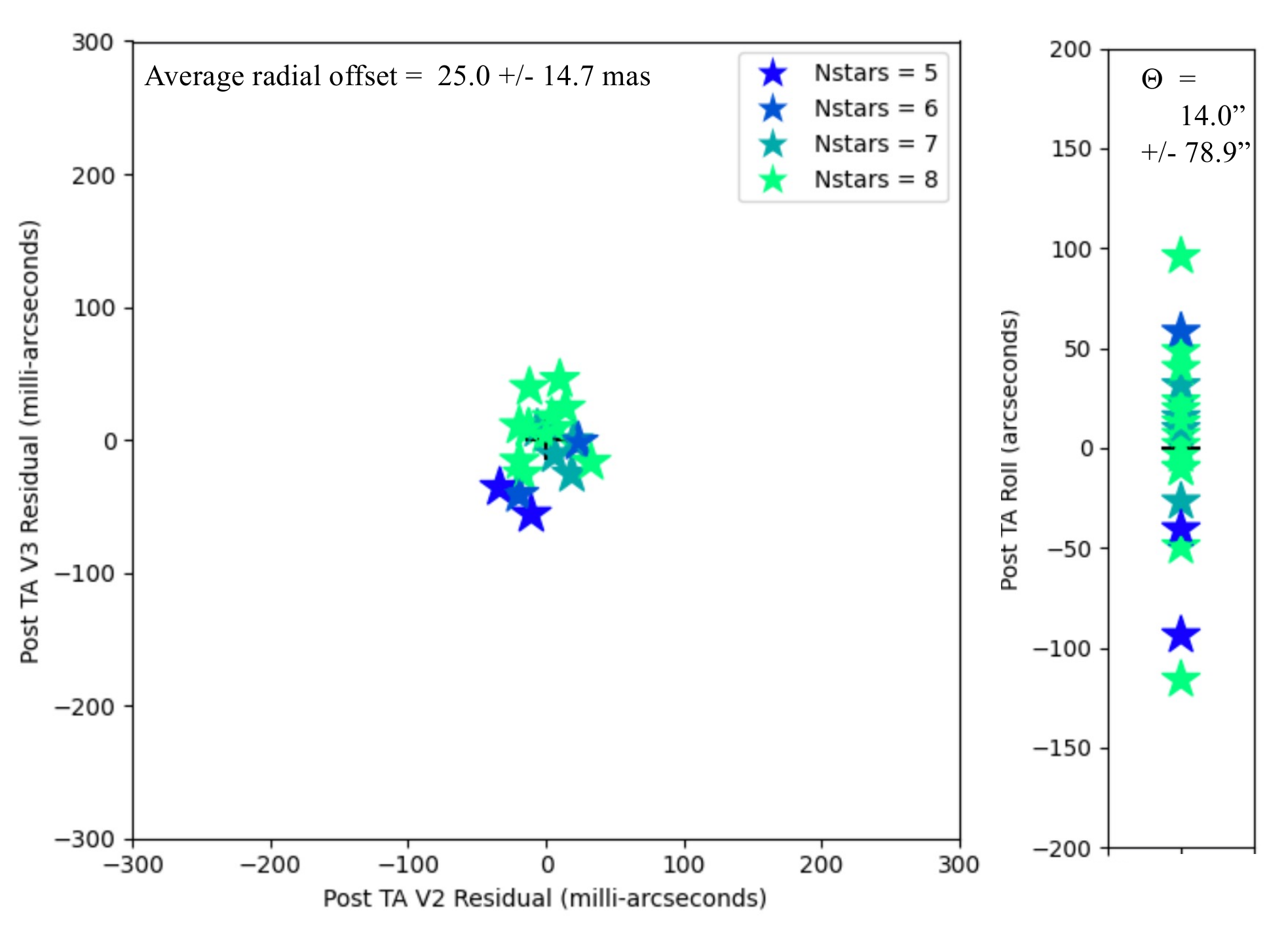}
\caption{Left: computed offset correction (in V2/V3 and roll angle) computed by the onboard algorithm for the ensemble of 20 MSATA visits analyzed so far. Each star symbol contains the average over the $n$ reference stars used for the computation (color-coded by the value of $5 \leq n \leq 8$). Right: the results after execution of the corrective slew computed via the onboard MSATA algorithm, as measured from the subsequent reference image (which still has the TA filter in place). }
    \label{fig:TAquality}
\end{figure*}

To evaluate the performance and post-TA target positioning accuracy, we have analysed the 20 NIRSpec MSATA procedures carried out so far. Figure\,\ref{fig:TAquality} shows their distribution of corrective TA slews (in V2, V3 and roll). We find all slew offset solutions to be within 100\,mas radial offset and at an average $-51\as$ roll offset compared to the optimal pointing. These relatively small corrections are a testament to the excellent 'blind' pointing accuracy of JWST which allows very accurate target placement even without any TA.

To illustrate the improvement achieved by the MSATA procedure, Fig.\,\ref{fig:TAquality} shows the target placement after MSATA execution for the same 20 observations. These measurements were  derived by running an offset analysis on the reference image that is acquired after the final TA slew is complete. As can be seen, the average radial offset over the ensemble of TA reference targets is 25.0\,mas, and the average roll offset from the optimal pointing solution is $14\as$. This is close to, but not yet fully in line with the requirements, but it should be kept in mind that the more critical displacement in cross-dispersion direction should be a factor of $\sqrt{2}$ smaller. In addition, many of the observations analyzed so far have used catalogs with non-optimal astrometric accuracy, and/or non-stellar objects as reference targets.

\section{NIRSpec Science Calibration Pipeline}\label{sec:pipeline}
NIRSpec data for all modes are processed by the JWST Science Calibration Pipeline\footnote{see https://jwst-docs.stsci.edu/jwst-science-calibration-pipeline-overview} in three stages. Stage 1 takes raw up-the-ramp integrations, applies various detector-level
corrections such as dark subtraction and linearity correction, flags jumps due to cosmic ray hits, and fits a slope to the ramp. Stage 2 calculates wavelength and spatial coordinates per pixel using the parametric instrument model, corrects for instrument throughput losses using the three-component flat field reference files, and converts to physical flux units (MJy/pixel for point sources, MJy/steradian for extended sources). The final stage combines
multiple exposures (such as from a nod or dither pattern) at the 2D level. High-level pipeline products include 1D and 2D calibrated spectra for FS and MOS modes and 3D data cubes for IFS mode.

The JWST Science Calibration Pipeline is a continual work in progress. The code is updated via four regular build releases per year with bug fixes and algorithm enhancements (although development versions are available on a continuous basis). Reference files are updated whenever new calibration data are available and indicate evolution in detector or throughput performance. As of the time of writing, there are some outstanding issues that users may want to be aware of:

i) the signal produced by snowball events is only partially corrected by the cosmic ray jump detection step. An optional algorithm that flags additional pixels affected by a snowball has recently become available, and further enhancements are being investigated.

ii) the `fast' throughput correction vector used during the S-flat step must be normalized by the wavelength interval falling onto a given detector pixel. The current reference files are normalized using representative values, which do not take into account the variation over the FOV. Tests are ongoing, but preliminary investigations suggest that this simplification adds an additional systematic uncertainty to the flux calibration that can be as high as ten percent. A future version of the pipeline will include an on-the-fly pixel-by-pixel normalization that should remove this source of uncertainty.

iii) in earlier versions of the pipeline, the resampling algorithms used in creating IFS cubes have sometimes produced wavelength-dependent artifacts in certain cases. A recent bug fix seems to fix the issue at least for point sources, but users should carefully inspect the extracted spectra to look for any unexpected behavior. Also note that because of the optical field distortion in the NIRSpec camera optics \citep[see][for a more detailed explanation]{jakobsen22}, all NIRSpec spectra are curved relative to the detector pixel grid, which in the case of compact or point sources leads to aliasing effects, esp. in the case of the gratings. Therefore, the spectrum of a single cube spaxel will show a sinusoidal variation in the extracted 1D spectrum. When summing up over a sufficiently large circular aperture, however, this effect should average out. For the same reason, the aliasing should be much reduced for extended sources. Ultimately, though, it is important to use dithering whenever possible, as the combination of dithered exposures will minimize such resampling artifacts.

iv) the 1D extraction apertures are automatically centered at the expected position of the source, based on sky coordinates specified in the JWST Astronomer's Proposal Tools (APT). For reasons that are still under investigation, this centering is often offset from the true position of the spectral trace, requiring manual specification of the center location by users.

v) there are no aperture corrections yet available for 1D extractions.
The default extraction apertures are set to be consistent with
those used for the generation of the F-flat reference files. For different aperture widths, users must currently compute their own aperture corrections using available observations of point sources.

vi) the outlier detection step in stage 3 of the pipeline is intended to catch any outliers missed by stage 1 by means of comparing the multiple input exposures. However, testing to date has shown performance problems when using the default thresholds. Users will need to adjust the parameters to get a reasonable identification of true outliers, although in many cases it may be preferable to turn the step off entirely until the algorithm is improved.

\section{Summary and Conclusions}
We have provided an overview of the NIRSpec performance as derived during the JWST commissioning campaign and the first few months of in-orbit operations. From a hardware perspective, the NIRSpec instrument performs nominally in all electrical, mechanical, and thermal aspects. When combined with the outstanding optical performance of the JWST OTE, and the better than expected accuracy and stability of Webb's Attitude Control System, it is no surprise that the scientific performance of NIRSpec  meets or exceeds the pre-launch expectations for all observing modes.

More specifically, we have shown that the complex on-board target acquisition procedures work well, the measured sensitivities are excellent across the board, and unrivalled by any other NIR spectrograph, and the photometric stability is very good which is important for time-series observations of exoplanets.

In terms of the calibration accuracy of NIRSpec science data, we have shown that the data acquired and the methods used so far allow the creation of reference files that are sufficient to meet the requirements for the photometric and wavelength calibration of NIRSpec spectra.

On the other hand, we have mentioned a number of open issues with the NIRSpec Science Calibration Pipeline, in particular for the flux calibration of NIRSpec spectra, the cube-building step for IFS data, and 1D spectral extraction. These issues are actively being worked, and improvements to the quality of the higher-level NIRSpec data in the Mikulski Archive for Space Telescopes (MAST) can be expected soon.

There are a few important `lessons learned' from these results that NIRSpec users should take note of:
\begin{itemize}[leftmargin=*]
\item 
the higher than expected PCE measured over most of the NIRSpec wavelength range may cause some exposures to saturate earlier than predicted from pre-launch estimates. This may make some `bright target' science less feasible and/or require modifications to the observing strategies, especially the selection of MSATA reference stars.
\item
 for MOS mode and the associated MSATA planning, it is essential that the catalog of MSATA reference stars is properly registered to the Gaia frame, not just in RA and Dec, but also in rotation.
 \item
 the initial telescope pointing after guide star acquisition is accurate enough that for most IFS observations, foregoing TA (by using TA=NONE or VERIFY\_ONLY) is feasible. Because of the better than expected PSF quality and the resulting small slit losses in the S1600A1 aperture, this may even be true for some time-series observations.
\end{itemize}

To summarize, the NIRSpec instrument onboard JWST presents a quantum leap for NIR spectroscopy in terms of sensitivity and multiplexing capability, and promises to have a lasting impact on many research fields.

\begin{acknowledgements}
We are deeply grateful to the large number of engineers and scientists in Europe, Canada, and the US, whose dedication and hard work over many years have turned NIRSpec and the entire JWST mission from a mere vision into reality.
\end{acknowledgements}

\bibliography{nirspec_refs}{}
\bibliographystyle{aasjournal}



\end{document}